\magnification1200

\vskip 2cm
\centerline {\bf 
Extended BMS representations  and strings}
\vskip 0.5cm

\centerline{Romain Ruzziconi ${}^1$ and Peter West ${}^2$}
\vskip 0.5cm
\centerline {${}^1${\it Center for the Fundamental Laws of Nature, Harvard University }}
\centerline { {\it 17 Oxford Street, Cambridge, MA 02138, USA} }

\centerline {${}^1${\it Black Hole Initiative, Harvard University,  20 Garden Street,  }}
\centerline {{\it Cambridge, MA 02138, USA}}
\centerline{}
\centerline{${}^2${\it Mathematical Institute, University of Oxford,}}
\centerline{{\it Woodstock Road, Oxford, OX2 6GG, UK}}
\centerline{${}^2${\it Department of Mathematics, King's College, London}}
\centerline{{\it The Strand, London WC2R 2LS, UK}}
\vskip 1cm
\centerline{ romainruzziconi@fas.harvard.edu\ \ \  peter.west540@gmail.com}
\vskip 2cm
\centerline {\bf Abstract}  
We construct in detail the irreducible representations of the BMS group with super rotations in three and four dimensions that have the same rest frame momenta as the massive and massless Poincar\'e point particles. We compare these representations to those of the Poincar\'e group and also to the analogous representations of global BMS. We argue that these extended BMS representations are carried by a string rather than a point particle. The super rotations play a crucial role in our discussions. 
\par 
\vskip2cm
\noindent

\vskip .5cm

\vfill
\eject

{\bf 1 Introduction}

A remarkable feature about general relativity is that, even far from the source of the gravitational field, the theory does not reduce to   
special relativity. This fact is particularly transparent in the context of asymptotically flat spacetime, where the asymptotic symmetry 
group is given by the Bondi,  van der Burg,  Metzner and  Sachs (BMS) group [1,2], and not simply the Poincar\'e.
group. It was subsequently  enhanced to include super rotations  [3,4,5,6],  this is called extended BMS. Later the symmetry was further enhanced by incorporating  diffeomorphisms on the celestial sphere  [7,8,9,10,11,44]. 

Although the importance of super rotations for scattering in flat space was understood through their relation with subleading soft graviton theorem [12,45], several fundamental questions remain. One of them concerns understanding the precise role of 
the BMS representations in formulating a well-defined theory of scattering amplitudes, with the long-term goal of better understanding 
the infrared divergences of the S-matrix as well as establishing  a holographic description of asymptotically flat spacetime. 

The  works of McCarthy [13,14,15,16,17] provided a mathematical classification of the unitary irreducible representations of the BMS group in four dimensions in the absence of the super rotations. A similar analysis was performed for the BMS group in three dimensions [18,19,20] including the super rotations.  The irreducible representations of $BMS_3$ were also discussed in reference [46] from the viewpoint of ultra relativistic limit of SL(2,R) modules. More recently the results of McCarthy were further discussed [21,22] in relation to the infrared structure of the S-matrix. Despite these results, this mathematical machinery has not yet found concrete  applications in the context of scattering theory and the role of super rotations in the representation theory remains unexplored. 

To fill this gap, we take a very pedestrian approach to constructing the wave functions associated with the irreducible representations of the extended BMS group in three and four dimensions that correspond to the  massive and massless representations of the Poincar\'e group. Such representations are of importance for the Carrollian proposal for flat space holography  [23,5,25,26,27,28,29] which hopes to determine the scattering of the usual particles by using BMS symmetry at infinity.  
\par
Our concrete approach consists in extending the standard Wigner construction of Poincar\'e particles [30] to the  infinite-dimensional extended BMS group which has the same semi-direct product structure. In particular we find the wavefunctions in super momentum space and Fourier transform them to find the wavefunctions in position space. We argue that these irreducible representations are carried by extended objects rather than particles and in particular strings. Unlike in the Poincar\'e case, taking the Fourier transformation requires an infinite number of  position  coordinates as we need one for each  momentum, including the super momentum.  These infinite coordinates can be interpreted as modes of a string that lives in the familiar spacetime. The super rotations are crucial for this conclusion as without them the super momenta  obey and infinite number of constraints which reduce their number to be that of the usual spacetime. We also discuss the form of the wavefunction in position space as we take the limit to time-like infinity  for the massive extended BMS irreducible representations in three dimensions. 
\par
The paper is organized as follows. In section two we find the irreducible representations of the extended BMS group in three dimensions  that correspond to the massive and massless particles of the Poincar\'e group.
We compare our results with those of references [18,19,20]. In section three we push the massive irreducible representations of  $BMS_3$  to time-like infinity $i^+$. In this section and section two we argue that the representations of extended $BMS_3$ are carried by a string.   
In section four we construct the irreducible representations of the extended $BMS_4$ group, including the super rotations,  in four dimensions that correspond to the massive and massless Poincar\'e particles.  In section five we  construct the irreducible representations of the BMS groups, without including the super rotations,  in three and  four dimensions that correspond to the massive and massless Poincar\'e particles. We compare our results with those of McCarthy. In section six we summarise and discuss our results. 
\par
The contents of this paper are as follows: 
\medskip
1 Introduction
\medskip
2 Irreducible representations of $BMS_3$
\medskip
\ \ \ 2.1 Massive $BMS_3$ representations

\ \ \ 2.2 Massless $BMS_3$ representations  
\medskip
3 Irreducible representations of Poincar\' e and $BMS_3$ at time-like infinity
\medskip
\ \ \ 3.1 Massive Poincar\' e particles at time-like infinity

\ \ \ 3.2 Massive $BMS_3$ irreducible representations at time-like infinity
\medskip
4 Extended $BMS_4$ irreducible representations
\medskip
\ \ \ 4.1 Massive extended $BMS_4$ representations

\ \ \ 4.2 Massless extended $BMS_4$ representations

\ \ \ 4.3 Interpretation of the extended $BMS_4$ representations
\medskip
5 Irreducible representations of BMS without the super rotations
\medskip
6 Summary and Discussion

\medskip
{\bf 2 Irreducible representations of  $BMS_3$  }
\medskip
In this section, we construct the irreducible representations of  $BMS_3$ which have the same rest frame momentum as those of the irreducible representations of the Poincar\'e group. 
The BMS$_3$ algebra [31,33] is given by
$$
[J_m, J_n] = i (m-n) J_{m+n} + i {Z_1\over 12} m (m^2- 1) \delta_{n+m,0}, \quad
$$
 $$
 [J_m, P_n] = i (m-n) P_{m+n} + i {Z_2 \over 12} m (m^2- 1) \delta_{n+m,0}, \quad   [P_m, P_n] = 0
\eqno(2.0.1)
$$ 
where $J_n$ and $P_n$, $n=0,\pm 1, \pm 2, \ldots $ are the super rotations and the super translations,  respectively and  $Z_1$ and $Z_2$ are the possible central generators. All the generators in the above commutators are related by a minus sign to the commutators in equation (5.7) given in  reference [18]. The values of the central charges for  three dimensional gravity in asymptotically flat spacetimes is given by $c_1=0$ and $c_2={3\over G}=-\langle Z_2 \rangle$.  
\par
The Poincar\'e subalgebra  in three dimensions has the generators ${\bf J}_{\mu\nu}$ and ${\bf P}_\mu$, $\mu,\nu=0,1,2$ which are contained in the  $BMS_3$ generators  $J_{-1}$, $J_0$, $J_1$, $P_{-1}$, $P_0$, $P_1$. The precise identification is given by  
$$
    J_0 = - {\bf J}_{12}, \quad {J}_{\pm 1} =  {\bf J}_{01} \pm i  {\bf J}_{02}, \quad  {\bf P}_0 =  {P}_0, \quad { P}_{\pm 1} =  {\bf P}_2 \mp i  {\bf P}_1
\eqno(2.0.2)$$ 
We take the following conventions for the Poincar\'e algebra 
$$
[{\bf J}_{\mu\nu}, {\bf J}_{\rho\kappa}]= \eta _{\nu\rho} {\bf J}_{\mu\kappa}-  \eta _{\mu\rho} {\bf J}_{\nu\kappa}
-   \eta _{\nu\kappa} {\bf J}_{\mu\rho}    +  \eta _{\mu\kappa} {\bf J}_{\nu\rho}, \ \ 
[{\bf J}_{\mu\nu}, {\bf P}_{\rho} ]= \eta _{\nu\rho} {\bf P}_{\mu}-\eta _{\mu\rho} {\bf P}_{\nu}
 \eqno(2.0.3)$$ 
\par
Since the super translations $P_n$ commute we can, just as in the case of the Poincar\'e algebra,  take the representations of the algebra to act on super momentum eigenstates $|p_n\rangle$ with $P_n |p_n \rangle= p_n |p_n\rangle$. Acting with a super rotation 
 the momentum eigenstates transform as $ |p^\prime_n\rangle= e^{\sum_m \Lambda_{-m} J_m} |p_n\rangle$. As a result we find that 
$$
P_n  |p^\prime_n\rangle= e^{\sum_m \Lambda_{-m} J_m} e^{-\sum_m \Lambda_{-m} J_m} P_n e^{\sum_m \Lambda_{-m} J_m} | p_n\rangle
\eqno(2.0.4)$$ 
Using the above commutators of  $BMS_3$ we find, for infinitesimal $\Lambda_m$, that the super momenta transform as 
$$
\delta p_n= -i \sum_m\Lambda_{-m} (-n+m) p_{n+m}-{ic_2\over 12} n(n^2-1)\Lambda _{n}
\eqno(2.0.5)$$
\par
The  $BMS_3$ algebra  is of the form of a semi-direct product of the algebra of the super rotations, $J_n$,  with the abelian super translations, $P_n$. As such it has the same generic form as the Poincar\'e algebra, which it contains,  and as such we can apply the Wigner method to find the irreducible representations. The irreducible representation of BMS$_3$ were previously studied in [18,19,20] and we  will discuss the relation between our  results and those given in that  paper. 
\par
The first step in the Wigner method is to choose which momenta  are non-zero in the rest frame and then find the isotropy group that preserves this choice. Since there are an infinite number of such generators there is a considerable choice. In this paper we will confine our attention to when only the usual momenta ${\bf p}_\mu$ are non-zero. As such we will make contact with the usual irreducible representation of the Poincar\'e group. We begin with the massive representation. 
\medskip
{\bf 2.1 Massive   $BMS_3$ representations}
\medskip
We choose the only non-zero rest frame super momentum  to take the values   $p_n^{(0)}$ to be $p_0^{(0)}= m= {\bf p }_0^{(0)}$, $p_n^{(0)}=0 , |n| \ge 1$.   This  corresponds to the usual massive particle with ${\bf p}^{(0)}_0=m, {\bf p}^{(0)}_1=0={\bf p}^{(0)}_2 $.   
To begin with we will take the central charge $c_2=0=c_1$ but we will consider $c_2$ to be  non-zero later on which is the case relevant for gravity in asymptotically flat spacetime. 
Using equation (2.0.5)  we find that this choice of momentum is preserved if 
$$
{ \delta p_n^{(0)}\over m}= 2in  \Lambda _n =0
\eqno(2.1.1)$$
Hence we find that  the group which preserves our chosen momentum $p_n^{(0)}$  is just  ${\cal H}^3_{m\not= 0}= \{J_0\}$ which is the same isotropy group as for the Poincar\'e case, namely SO(2)  as  $J_0= J_{12}$. 
\par
Next we take an irreducible  representation of the isotropy group ${\cal H}^3_{m\not=0}$  carried by the  states $| p_n^{(0)},a \rangle$ 
which obey $P_n |  p_n^{(0)},a \rangle=  m \delta_{n,0}  |  {p_n^{(0)}} , a \rangle$ where  the index transforms under ${\cal H}^3_{m\not=0}= \{J_0\}$. For example, for  spin one this representation is given by 
$$
J_0 |  p_n^{(0)},1 \rangle= |  p_n^{(0)},2 \rangle, \quad J_0 |  p_n^{(0)},2 \rangle=- |  p_n^{(0)},1 \rangle
\eqno(2.1.2)$$
\par
The general state in the irreducible representation is given by 
$$
|  p_n,a \rangle \equiv e^{\sum_{|n|\ge 2} \varphi_{-n} J_n} e^{\varphi_{-1} J_{1}+ \varphi_{1} J_{-1}} |  p_n^{(0)},a \rangle
\eqno(2.1.3)$$
We will refer to this representation, which is based on the above choice of rest frame momentum, as the massive irreducible representation of  $BMS_3$.  Similarly in the next section  we will refer to the massless   $BMS_3$ irreducible representations as the one that  corresponds to the same choice of momenta as for the massless irreducible representation of the Poincar\'e group. 
\par
By acting on the state $|  p_n,a \rangle $ with $P_n$ we find its  eigenvalue $p_n$, indeed 
$$
{ p_n\over m}= \delta_{n,0}  +2in\varphi _{n}+\sum_p\varphi_{p+n}\varphi_{-p} (p^2-n^2)+\dots 
\eqno(2.1.4)$$
We recall that $\varphi_0=0$ and in deriving this equation we have combined the exponentials in equation (2.1.3) and the $\varphi_n$ in the last equation are the new variables. This irreducible representation is in the list given in [18].
\par
It is instructive to consider the restriction of the above to the Poincar\'e case where the general state is given by 
$$
|  p_n, a \rangle \equiv e^{\varphi_{-1} J_1+ \varphi_{+1}J_{-1}} |  p_n^{(0)},a \rangle= e^{(\varphi_{+1}+\varphi_{-1}) J_{01} -i(\varphi_{+1}-\varphi_{-1}) J_{02}} |  p_n^{(0)},a \rangle
\eqno(2.1.5)$$
We find at lowest level 
$$
{p_0\over m}=  1 +2\varphi_{-1} \varphi_{+1} +\ldots, \quad 
{p_{+1} \over m}=  { 2}i \varphi_{+1} +\ldots , \quad 
$$
$$
{ p_{-1}\over m} =  -{ 2}i \varphi_{-1} +\ldots
 \eqno(2.1.6)$$
The momenta are parameterised by the two degrees of freedom  $\varphi_{\pm 1}$ which is the correct count given that ${\bf p}_\mu {\bf p}^\mu=- p_0^2+ p_{1}p_{-1}=-m^2$. 
\par
In momentum space the wavefunction of equation (2.1.5) is a function of $\varphi_{\pm 1}$,  or equivalently ${\bf p}_\mu, \mu=0,1,2$ subject to ${\bf p}_\mu {\bf p}^\mu=-m^2$. 
We denote the Cartesian coordinates of spacetime to be given by $X^\mu$, $\mu=0,1,2$ and the    Fourier transform to x space is given by
$$
\Psi (X^\mu,a) = \int {d^{3} {\bf p} \over (2\pi)^{3}} \delta ( {\bf p}^\mu {\bf p}_\mu + m^2 )  e^{i{\bf p}^\mu X_\mu} \Theta ({\bf p}^0) \psi ({\bf p},a)=  \int {d^{2} {\bf p} \over (2\pi)^{2}{\bf p}^0}   e^{i{\bf p}^\mu X_\mu}  \psi ({\bf p}, a)
$$
$$
= \int {d\varphi_1\varphi_{-1} \over (2\pi)^2 2 } {\sinh \varphi (\varphi_1-\varphi_{-1})^2\over \varphi^3}
e^{i(-x_0p_0 + x_{1} p_{-1} + x_{-1} p_{1})}  \psi ({ \bf p},a) 
$$
$$
= \int {d\varphi_1\varphi_{-1} \over (2\pi)^2 2 } {\sinh \varphi (\varphi_1-\varphi_{-1})^2\over \varphi^3}
e^{i(x_0p_0 -2ix_{1}\varphi_{-1} +2i x_{-1} \varphi_{1})}  \psi ({\bf  p},a) 
\eqno(2.1.7)$$
where $\psi(p,a)= \langle p | p_n,a \rangle$.  In going between the first  and second lines we have identified $x_0$ and $x_{\pm 1}$ to be given by 
$$
x_0= X_0 , \quad x_{\pm 1}\equiv {1\over {2}}(X^2 \mp iX^1)
\eqno(2.1.8)$$ 
Hence we find the usual result, a  function $\Psi(x) $ subject to $(\partial^\mu \partial_\mu -m^2)\Psi=0$. 
\par
The situation for  $BMS_3$ is rather different. Equation (2.1.3) shows that the wavefunction in momentum space depends on $\varphi_n, \ n\not= 0$, or equivalently $p_n$ for all $n$,  subject to  the  condition 
$ (-P_0^2+ P_{1} P_{-1}+m^2) |p^{(0)}_n,a \rangle=0$ . As a result we find that 
$$
e^{\sum_{|n|\ge 2} \varphi_{-n} J_n} e^{\varphi_{-1} J_{1}+ \varphi_{1} J_{-1}}(- P_0^2+ { P}_{1} { P}_{-1}+m^2)
e^{-(\varphi_{-1} J_{1}+ \varphi_{1} J_{-1})}e^{-\sum_{|n|\ge 2} \varphi_{-n} J_n} 
$$
$$
e^{\sum_{|n|\ge 2} \varphi_{-n} J_n} e^{\varphi_{-1} J_{1}+ \varphi_{1} J_{-1}} |p^{(0)}_n \rangle
=C(p_n) |p_n,a \rangle=0
\eqno(2.1.9)$$
 where 
  $$
c(P_n)\equiv e^{\sum_{|n|\ge 2} \varphi_{-n} J_n} e^{\varphi_{-1} J_{1}+ \varphi_{1} J_{-1}}(-{ P}_0^2+ { P}_{1} { P}_{-1}+m^2)
e^{-(\varphi_{-1} J_{1}+ \varphi_{1} J_{-1})}e^{-\sum_{|n|\ge 2} \varphi_{-n} J_n} 
$$
$$
= -{ P}_0^2+ { P}_{1} { P}_{-1}+m^2+\sum_{n , |n|\ge 2} i\varphi_{-n}(-2n { P}_{n}{ P}_0 +(n-1)  { P}_{n+1}{ P}_{-1} +(n+1)  { P}_{n-1}{ P}_{1}+m^2 )+\ldots 
$$
$$
=-{ P}_0^2+ { P}_{1} { P}_{-1}+m^2+\sum_{n , |n|\ge 2}{{ P}_{-n} \over 2n}(2n { P}_{n}{ P}_0 -(n-1)  { P}_{n+1}{ P}_{-1} -(n+1)  { P}_{n-1}{ P}_{1} )+\ldots 
\eqno(2.1.10)$$
where we have used equation (2.1.4) and $+\ldots$ means terms of order $P^4$ and higher. At very lowest order $p_0=m+\ldots $ and so at lowest order $C(p)= -p_0^2+\sum_{n,n\not=0}p_np_{-n}+m^2=0$.
Thus the super momenta of the states obey  the  constraint $C(p_n) =0$ and hence  the wavefunction in momentum space  depends on an infinite number of variables. We note that the above general states of equation (2.1.3) do not include those with only the usual massless momentum of the Poincar\' e group, that is,  they cannot have ${\bf p}^\mu {\bf p}_\mu=0$ with all higher super momenta being zero. This justifies our division of irreducible representations into massive and massless. 
 \par
We can take the Fourier transform to x space for the above massive  $BMS_3$ irreducible representation. However,  now we must introduce a space that is parameterised by an infinite number of variables which we denote by $x_n, \ n=0,\pm 1,\pm2 ,\ldots $  as we need one coordinate for each  momentum $p_n$. 
The Fourier transform has the  generic form 
$$
\Psi (x_n,a)= \int \prod _n d\varphi_n J e^{i\sum _n p_n x_{-n} } \psi ({p_n},a ) 
\eqno(2.1.11)$$
where $J$ is a function of $\varphi_n$ which is required for an invariant measure. The wavefunction $\Psi(x_n,a)$ is subject to one equation, which is the x-space version of equation (2.1.9), namely $C(-i{\partial\over \partial_{x_n}} )\Psi(x_n,a)=0$.
\par
We choose the  $x_n$ to be coordinates on the coset space constructed from   $BMS_3$ with the subgroup being the group generated by the super rotations $J_n$. This mimics the case for the Poincar\'e group where the usual spacetime is the coset space of the Poincar\'e group divided by the Lorentz group. The coordinates $x_n$ then transform as 
$$
\sum_n x_{-n}^\prime P_n= e^{-\sum_m \Lambda_{-m} J_m} \sum_n x_{-n} P_n e^{\sum_m \Lambda_{-m} J_m}  \Rightarrow 
\delta x_n=- i \sum_m (2m+n) \Lambda_{-m} x_{n+m}
\eqno(2.1.12)$$
\par
At first sight it looks as if we have a particle moving in an infinite spacetime but this is not consistent with the fact that the  $BMS_3$ algebra is derived in three dimensional spacetime. To better understand the meaning of the many coordinates we will reformulate the above by introducing $z=e^{i\theta}$ and the quantities 
$$
J(z)=\sum_m J_m z^{m} , \ \  \Lambda(z) =\sum_m \Lambda_m z^{m} , \ \ X(z)= \sum_m x_m z^{m}
 \eqno(2.1.13)$$
 In which case we can write the transformation of the coordinates of  equation (2.1.12) as  
$$
\delta X(z) =-i \Lambda(z) z{dX(z) \over dz} +i z{d\Lambda(z) \over d z} X(z)
\eqno(2.1.14)$$
Thus $X(z)$ transforms under diffeomorphisms of $z$ as a weighted scalar which is consistent with interpreting $z$ as a parameter labelling a one dimensional object. We can then write the   wavefunction in x-space as $\Psi(X(z),a)$. It follows that the massive irreducible representation of  $BMS_3$ should be viewed as  an  object of dimension one, that is, a string which is    parameterised by $z$. We will discuss this string interpretation  in greater detail in section 3.2. 
\par
We now find the massive irreducible representation when the central charge $c_2\not=0$. In this case the momenta vary under the super rotations as given in equation (2.0.5). 
Taking into account that the rest frame momentum is given by  $ p_n^{(0)} =m\delta_{n,0}$ we find that the condition to preserve this choice of momentum is given by 
$$
{\delta p_n  \over m}= ni(2 -{c_2\over 12} (n^2-1))\Lambda_{n}=0
\eqno(2.1.15)$$
In the generic case, the isotropy group is the same as above, namely the isotropy group just contains the generator $J_0$. However, if the central charge is such that $n =  \sqrt{1 + 24/c_2}$ is an integer, then the isotropy group is ${\cal H}_{mass}^c=\{ J_0, J_n, J_{-n} \}$ which is the algebra of SL(2,R). These irreducible representations are also in the list of [18].  There are only a limited number of solutions if $c_2$ is positive as it should be. We note that if $c_2$ is negative then there is only one solution, that is, $c_2=-24$ with $n=0$ but this does not lead to a larger isotropy group. The situation is a bit like that for supersymmetry where for  certain values of the central charge the irreducible representations of supersymmetry are smaller. 
\par
If the central charge does not have one of these special values the general momentum state is given by 
$$
|  p_n,a \rangle \equiv e^{\sum_{|n|\ge 1} \varphi_{-n} J_n}  |  p_n^{(0)},a \rangle
\eqno(2.1.16)$$
While if $n = \pm \sqrt{1 + 24/c_2}$ is an integer  then we have the above state but must set $\varphi_n=0=\varphi_{-n}$. 
\par
The momenta of this state are given in terms of the $\varphi_n$ by 
$$
{p_n\over m}= \delta_{n,0} + 2in\varphi_n (1-{c_2\over 24}(n^2-1))+ \sum_p (p^2-n^2) \varphi_{-p} \varphi_{n+p}((1-{c_2\over 24}((n+p)^2-1))+\ldots 
\eqno(2.1.17)$$
We note that $-p_0^2+\sum_n p_np_{-n}+\dots =0$ where $+\ldots $ indicates higher terms in powers of $\varphi_n$ and the central charge. 
\par
To construct the wavefunction in x space we must introduce a coordinate for each independent momenta and so we introduce $x_n$, $n\ge 1$ as well as a coordinate $y$ corresponding to the presence of the central charge generator $Z_2$. The transformation of these coordinates follows from demanding that 
$$
\sum_n x_{-n}^\prime P_n+ y^\prime Z_2= e^{-\sum_m \Lambda_{-m} J_m} (\sum_n  x_{-n} P_n +yZ_2)e^{\sum_m \Lambda_{-m} J_m}  
\eqno(2.1.18)$$ 
We find that 
$$
\delta x_n=- i \sum_m (2m+n) \Lambda_{-m} x_{n+m} ,\ \  \delta y=i \sum_n x_n \Lambda _n n(n^2-1) 
\eqno(2.1.19)$$
We can rewrite this in terms of the quantities of equation (2.1.14) as 
$$
\delta X(z) =-i \Lambda(z) z{dX(z) \over dz} +i z{d\Lambda(z) \over d z} X(z) , 
\ \ \delta y= i\int dz z\Lambda (z) {d^3\over dz^3 }(zX(z))
\eqno(2.1.20)$$
\par
The wavefunction in x space then has the form $\Psi( X(z), y,a)$. The dependence on the coordinate $y$ could be rather trivial and it may be related to the central charge the string may carry.


\medskip
{\bf 2.2 Massless extended $BMS_3$ representations}
\medskip
We choose the rest frame momentum to be the  momentum one usually takes for the massless irreducible representation of  the Poincar\'e group, namely  ${\bf p}^{(0)+}\equiv  {1\over \sqrt 2}({\bf p}^{{(0)2}}+{\bf p}^{{(0)}0})=1$, ${\bf p}^{{(0)}-}\equiv {1\over \sqrt 2}({\bf p}^{{(0)}2}-{\bf p}^{{(0)}0})=0$ and ${\bf p}^{{(0)}1}=0$, which has ${\bf p}^\mu {\bf p}_\mu =0$,  and all other super momenta vanishing. This is equivalent to 
$$
p_0^{(0)}=-{1\over \sqrt 2}, \ p_1^{(0)}=p_{-1}^{(0)}={1\over \sqrt 2} , \quad p^{(0)}_n=0 , |n|\ge 2
\eqno(2.2.1)$$
The first step is to find the isotropy  group, which preserves this  momentum. The variation of the super momenta $P_n$ under the super rotations  is given in equation (2.0.5) and so the super rotations preserving this  choice of super momenta  satisfy 
$$
\delta p_n^{(0)}=0= -{i\over \sqrt 2}(2n\Lambda _{n} (1+{\sqrt {2} c_2\over 24}(n^2-1))+(1-2n) \Lambda _{n-1} -(1+2n) \Lambda _{1+n})
\eqno(2.2.2)$$
At lowest orders we find the conditions on the parameters 
$$
 \quad \Lambda_1 - \Lambda_{-1} = 0 \ , \quad 
  2 \Lambda_{1} - \Lambda_0 - 3 \Lambda_{2} = 0  , \quad 
    2 \Lambda_{-1} - \Lambda_0- 3 \Lambda_{-2}  = 0  , \quad 
$$
$$
   \quad 4 \Lambda_{2} (1+{\sqrt{2} c_2\over 8}) - 3 \Lambda_{1} - 5 \Lambda_{3} = 0  , \quad
    \quad 4 \Lambda_{-2} (1+{\sqrt{2} c_2\over 8}) - 3 \Lambda_{-1} - 5 \Lambda_{-3} = 0  , \ldots 
\eqno(2.2.3)$$   
Taking $n\to -n$ in equation (2.2.2) it is easy to realise that $\Lambda_n=\Lambda_{-n}$. 
 In the BMS$_3$ case we find that  equations (2.2.3) can be solved in terms of two real parameters $a$ and $b$ as follows 
$$
 \Lambda_{1} = a = \Lambda_{-1}, \quad \Lambda_0 = 2b, \quad \Lambda_{2} = {2\over 3} (a-b) = \Lambda_{-2},$$ 
$$
 \Lambda_{3} =  - {{(a+ 8 b-\sqrt{2} c_2(a-b))}\over{15}} = \Lambda_{-3}, \quad \ldots
 \eqno(2.2.4)
$$
As a result the isotropy group for the massless $BMS_3$ is 
${\cal H}_{m=0}^3= \{H_1, H_2\}$  where 
$$
H_1= J_0 -{1\over 3}(J_2+J_{-2}) -{4\over 15}(1+{\sqrt{2} c_2\over 8}) (J_3+J_{-3})+\ldots ,\quad
$$
$$
H_2= (J_{1}+ J_{-1})+{2\over 3}(J_2+J_{-2}) -{1\over 15}(1-\sqrt{2} c_2)(J_3+J_{-3})+\ldots
\eqno(2.2.5)$$
As such the isotropy group has two elements and this conclusion appears to be independent of whether $c_2$ is zero or not. 
\par
It will be instructive, following [18], to rephrase the above steps in terms of the quantities. 
$$
  P(z) = \sum_m P_m z^m, \quad \Lambda (z) = \sum_n \Lambda_n z^{n}  , \quad J(z)=\sum_m J_m z^m 
 \eqno(2.2.6)$$
where $z = e^{i\theta}$.
Under a super rotation $\int {dz \over z} \Lambda (z) J(z)$ the super momentum transformation of equation (2.0.5) can be written as 
 [18,19,20]
$$
\delta  p(z) =  i z \Big(\Lambda(z)  {d\over dz} p(z) + 2  p(z) {d\over dz}\Lambda(z) -{c_2\over 12} {d^3\over dz^3} (z\Lambda(z))\Big) 
 \eqno(2.2.7)$$
Thus our rest frame momenta will be preserved if 
$$
\delta  p(z) =0=  iz  \Big(\Lambda(z)  {d\over dz} p^{(0)}(z)  + 2  p^{(0)} (z) {d\over dz}\Lambda(z) -{c_2\over 12} {d^3\over dz^3} (z\Lambda(z))
\Big) 
 \eqno(2.2.8)$$
which we can rewrite as 
$$
{d\over dz} (\Lambda (z)^2 p^{(0)}(z)-{c_2\over 12} {d^2\over dz^2} (z\Lambda(z))  = 0 
 \eqno(2.2.9)$$ 
\par
As we now illustrate one has to handle this equation with care. For simplicity we will illustrate this for $c_2=0$. In this case it  might 
 appear that one can rewrite equation (2.2.9) as  
$$
 \Lambda (z)^2 p^{(0)}(z) = \rm{e^2} \ \ {\rm or \ }\ \  \Lambda (z) \sqrt{ |p^{(0)}(z)|} =\pm \rm{e}
  \eqno(2.2.10)$$ 
where $e$ is a constant. 
\par
It is tempting to  conclude  that, if $p^{(0)}(z)$ does not have any zeros, then we can take $\Lambda (z) ={ {e}\over  \sqrt{|p^{(0)}(z)}|}$. In this case the isotropy group has  one element $\int {dz\over z} \Lambda (z) J(z)$ and so is a group of dimension one. This is the case of the massive particle as $p^{(0)}{(z)}=1$, and as a result   $\Lambda =\pm e$, and so  the isotropy group  contains the one generator $J_0$ in agreement with what we found in section (2.1). In the massless case $p^{(0)}(z) =- {1\over \sqrt{2}}( 1-( z + z^{-1}))$  which has two zeros as $z={1\over 2}(1\pm \sqrt {3}i)$. Taking this value in equation (2.2.10), we find that $e=0$ and so 
$\Lambda=0$. This implies  that the isotropy group has no elements which  is not in agreement with the above two dimensional isotropy group.  
\par
We note that 
$$
2^{{1\over 4}} \sqrt{   | p^{(0)}|} = (1-(z+z^{-1}))^{{1\over 2}}= 1 - {1\over 2} ( z + z^{-1}) - {1\over 4}  ( z + z^{-1})^2 -{3\over 8}  ( z + z^{-1})^3- {15\over 2^4} ( z + z^{-1})^4+\ldots 
$$
$$
=1-{1\over 2} -{90\over 2^3}+\dots + {\rm z \ dependent \ terms}
\eqno(2.2.11)$$ 
The problem is that this is not a convergent series as the constant term goes as $2^{2n}$ for large $n$.  As a result one cannot take the square root as in  equation (2.2.10).  This problem is generic for any rest frame momenta which, when written as $p^{(0)}(z)$,  possess positive and negative powers of $z$. 
\par
The $\Lambda (z)$ corresponding to the isotropy group of equation (2.2.5) is given by 
$$
\Lambda (z) = 2b +a (z^1+z^{-1}) +{2\over 3} (a-b) (z^2+z^{-2}) -{(a+8b)\over 15}  (z^3+z^{-3})+\ldots 
\eqno(2.2.12)$$
One can verify that this does satisfy equation (2.2.8)   at lowest orders when taking $p^{(0)}(z) = -{1\over \sqrt{2}}( 1-(z + z^{-1}))$.
In fact the same divergence problem occurs when we take $(\Lambda(z))^2$ and so one cannot reliably use equation (2.2.9), or indeed the first equation in (2.2.10). We note that equation (2.2.8) does  not involve any divergent expressions and so can be used to find the isotropy group. 
\par
It is instructive to consider the restriction of the above  to the Poincar\'e group. In this case $\Lambda_n=0 , \ |n| \ge 2$ 
and we only consider the action of $J_0$ and $J_{\pm 1}$ and $p_0$ and $p_{\pm 1}$.  Looking at the first three constraints of equation (2.2.3) we find that $\Lambda_0=2\Lambda_1= 2\Lambda_{-1}$ which is equivalent to taking $a=b$. As such the  isotropy group has the one generator $(J_0+{1\over 2} (J_{1}+J_{-1})=-{\sqrt 2} J_{1+}$ where $J_{1+}$ is the Lorentz generator in light-cone notation.  The remaining Lorentz generators are ${\sqrt 2} J_{1-}= (-J_0+{1\over 2}(J_{1}+J_{-1}))$ and $J_{+-}=J_{02}=  -{i\over 2} (J_1-J_{-1})$. In the massless  irreducible representation of the Poincar\'e group the generator $J_{1+}$ is trivially realised on the rest frame states and as such the effective isotropy group is trivial. As such the general state for the massless Poincar\'e irreducible representation  is given by 
$$
|{\bf p}_\mu \rangle = e^{\varphi J_{1-}}e^{\phi J_{+-}} |{\bf p}^{(0)}_\mu \rangle 
\eqno(2.2.13)$$
The boost in effect contains the generators $J_1$ and $J_{-1}$. 
\par
Returning to the $BMS_3$ irreducible representation, the  general state in the massless  irreducible representation of $BMS_3$ can be written as 
$$
| p_n\rangle= e^{\sum _{n , n\not= 0} \varphi_{-n} J_n} |  p_n^{(0)}, a \rangle= e^{\varphi_{-1} (J_1 - J_{-1}) + \sum _{|n|\ge 2}\varphi_{-n} J_n }|  p_n^{(0)}, a\rangle
\eqno(2.2.14)$$
where $\varphi_0=0$ and $\varphi_{1}=-\varphi_{-1}$. Even when restricted to just the Poincar\'e group, that is, $\varphi_n=0$ for $|n|\ge 2$, the massless extended $BMS_3$ general state is different to that for the Poincar\'e case in that  even restricted to these generators the latter  has  an additional boost  $J_1+J_{-1}$. 
\par
The momenta of the general state of equation (2.2.14) are given, when $c_2=0$,  by 
$$
p_m= p_m^{(0)}-{i\over \sqrt 2}(2m\varphi_{m} +(1-2m) \varphi_{m-1} - (1+2m) \varphi_{1+m} )
$$
$$
-{1\over 2\sqrt 2}\sum_{n} \varphi_{-n} \{(n^2-m^2) \varphi_{n+m}+ (n-m) (1-2(n+m)) \varphi_{n+m-1}
$$
$$
- (n-m) (1+2(n+m)) \varphi_{1+n+m})\}+\ldots 
\eqno(2.2.15)$$
subject to the above condition,  $\varphi_0=0$ and $\varphi_{1}=-\varphi_{-1}$. 
\par
In the Poincar\'e case we have the momentum ${\bf p}_\mu$ which is subject to the one condition ${\bf p}_\mu{\bf p}^\mu=0$ leaving  two independent components. Calculating the momentum of the general state of equation (2.2.13) we find that it is expressed in terms of $\varphi$ and $\phi$, which is in agreement.  In the massless  $BMS_3$ case we have an infinite number of momenta which can be expressed in terms of 
$\varphi_{-1}$ and $\varphi_n$, $|n|\ge 2$ but not $\varphi_0$ or $\varphi_1+\varphi_{-1}$ as these vanish. This  suggests that the super momenta are subject to two constraints, one of which will be the analogue of the Poincar\'e condition. 
\par
There is, however,  a problem with the above  isotropy group ${\cal H}^3_{m=0}$: computing the  commutator of the two generators appears to give  a divergent result.  One could take one of the generators to vanish on the rest frame state leaving the other to have a non-trivial action. In particular  one could take the latter generator to be the one with  $a=b$ which agrees with the  isotropy group for  the massless Poincar\'e group when restricted to this case. It could be that this step is required by taking the super momenta to satisfy certain topologies.  

\medskip
{\bf 3 Irreducible representations of Poincar\'e and  $BMS_3$ at time-like infinity}
\medskip
In this section we  push the massive representations of the Poincar\'e group in any dimension to time-like infinity $i^+$. For the massless Poincar\'e particle this was carried out in [34] and we follow this paper for the massive Poincar\'e  particles. 
We then carry out the analogous construction for the massive irreducible representation of $BMS_3$ to find it is described by a string. 

\medskip
{\bf 3.1 Massive Poincar\'e particles at time-like infinity}
\medskip
We will first  look at how the irreducible representations of the Poincar\'e group behave  at time-like infinity. This  has already been discussed in [35,47], but we will revisit it using the techniques of section 5 of reference [34] which were used to find the behaviour of the  massless irreducible representations of the Poincar\'e group at null infinity. We will carry out this calculation in a spacetime with $D$ dimensions. This approach uses  only group theory manipulations starting from the irreducible representations. 

Let us choose the rest frame  momentum to be given by 
$$
{\bf  p}^{(0)}_\mu = m (1, 0, \ldots, 0)
\eqno(3.1.1)$$ such that ${\bf p}^{(0)}_\mu {\bf p}^{{(0)}\mu} = -m^2$. The isotropy group is given  by 
$$
 {\cal H}_{m\not=0}^{Poincare} = \{ J_{ij} ,\quad i,j=1,\ldots ,D-1 \}
\eqno(3.1.2)$$ 
which form the ${SO}(D-1)$ algebra. The  irreducible representation of the Poincar\'e group is built from an irreducible representation of ${\cal H}_{m\not=0}^{Poincare}$  carried  by a wavefunction of the momentum ${\bf p}^{(0)}_\mu $, namely $ \psi ({\bf p}^{(0)},a)$, which  satisfies ${\bf P}_\mu \psi_ ({\bf p}^{(0)},a) = {\bf p}^{(0)}_\mu \psi ({\bf p}^{(0)},a)$. This wave function carries an index $a$ which transforms in an irreducible representation of $  {\cal H}_{m\not=0}^{Poincare} $. 
\par
 The wave function associated with a generic momentum ${\bf p}^\mu$ in the orbit of ${\bf p}^{(0)}$ is by definition  given by 
$$
 \psi ({\bf p},a) = e^{\sum_j \hat \varphi_j J_{j0}} \psi ({\bf p}^{(0)},a)
 \eqno(3.1.3)$$ 
and satisfies ${\bf P}_\mu \psi (p,a) = {\bf p}_\mu \psi ({\bf p},a)$. 
\par
The value of the momenta is determined by the action of  the momentum operator on the general state and we find that 
$$
 {\bf p}_\mu  = m (\cosh \hat \varphi ,  {\hat \varphi_i\over \hat \varphi} \sinh \hat \varphi ) , \quad  i=1,\ldots , D-1
\eqno(3.1.4)$$
where  $\hat \varphi= \sqrt {\sum_{i=1}^{D-1}\hat \varphi_i\hat \varphi_i}$. Rather than take the momentum to be  a function of the $D-1$ $\hat \varphi_i$ we will take it to be a function of $\eta_i= {\hat \varphi_i\over \hat \varphi} , i=1,\ldots , D-2$ and $\hat \varphi$. We note that $\eta^i \eta_i = 1$. 
\par
The wave function in position  space is given by 
$$
\Psi  (X,a) = \int {d^{D} {\bf p} \over (2\pi)^{D-1}} \delta ( {\bf p}^\mu {\bf p}_\mu + m^2 )  e^{i{\bf p}^\mu X_\mu} \Theta ({\bf p}^0) \psi  ({\bf p}, a)=  \int {d^{D-1} {\bf p} \over (2\pi)^{D-1}{\bf p}^0}   e^{i{\bf p}^\mu X_\mu}  \psi ({\bf p},a)
 \eqno(3.1.5)$$
 where the second integral is over the $D-1$ spatial components of the  momentum. Rather than integrate over the momenta we can integrate over $\eta_i$ and $\hat \varphi$. The corresponding  Jacobian $J$ is given by 
$$
J = {(\sinh \hat \varphi)^{D-2} \cosh \hat \varphi\over \eta_{D-1}}
\eqno(3.1.6)$$ 
where  it is understood that $(\eta_{D-1})^2 = 1- \sum_{i=1}^{D-2} \eta_i^2$.   As a result we find that 
$$
\Psi  (X,a) =
\int d\hat \varphi \prod_{i=1}^{D-2}  d \eta_i { (\sinh \hat \varphi)^{D-2} \over 2 (2\pi)^{D-1}\eta_{D-1}}  e^{i{\bf p}^\mu X_\mu} \psi  ({\bf p},a) 
\eqno(3.1.7)$$
\par
We now wish to push the associated position space wave function  to time-like infinity. We will focus on future time-like infinity $i^+$, but the analogue procedure can be repeated for past time-like infinity $i^-$. Let us introduce the spacetime coordinates $\tau$ and $y^\mu$ as follows 
$$ 
X^\mu = \tau y^\mu , \qquad y^\mu \eta_{\mu \nu} y^\nu = -1
\eqno(3.1.8)$$ In this parametrization, the flat space metric reads as
$$
 ds^2 = \eta_{\mu \nu} dX^\mu dX^\nu = - d\tau^2 + \tau^2 \eta_{\mu\nu} dy^\mu dy^\nu
\eqno(3.1.9)$$ 
Each $\tau = {\rm constant}$ leaf in the foliation is a hyperboloid, and taking $\tau \to + \infty$ leads us to $i^+$. A convenient  parametrization on the hyperboloid is given by 
$$
 y^\mu = (\cosh \zeta  , {\zeta^i \over \zeta}\sinh \zeta ),  \quad {\rm where} \quad \zeta = \sqrt {\sum_{i,j}^{D-1}\zeta^i \zeta^j\delta_{ij}}
\eqno(3.1.10)$$ 
with $i= 1, \ldots,D-1$. While one could take the variables to be $\zeta^i$, $i=1,\ldots , D-1$  in what follows we will take them to be 
$ \xi^i ={\zeta^i \over \zeta} $, $i=1,\ldots , D-2$  and $\zeta$. 
 In terms of these coordinates the  metric reads as
$$
ds^2 = - d\tau^2 + \tau^2 (d\xi^2 + \xi^2 \sinh^2 \zeta \ d\Omega^2_{D-2} )
\eqno(3.1.11)$$ 
where $d\Omega_{D-2}^2$ is the unit round sphere metric on $S^{D-2}$. 
\par
We want to consider the representation of a massive particle at time-like infinity   in equation (3.1.7). The argument of the exponential is given by
$$
{\bf p}^\mu X_\mu = \tau m g (\eta_i, \zeta,  \xi_i, \hat \varphi, \eta_i), \qquad  
$$
$$
g (\eta_i,  \zeta ,  \xi_i , \hat \varphi, \eta_i) = \cosh (\hat \varphi - \zeta ) + {1\over 2} \sum_{i=1}^{D-1} (\xi^i + \eta^i)^2 \sinh \zeta \sinh \hat \varphi 
\eqno(3.1.12)$$
We will now take the limit of $\tau  \to +\infty$ of $\Psi_\sigma (X)$ to obtain the states at the boundary $i^+$ by using the stationary phase approximation formula 
$$
 \lim_{\tau\to \infty} \int _a^b dx f(x) e^{i\tau g(x) }= \lim_{\tau\to \infty} f(z) e^{i\tau g(z)} \sqrt{{2\pi i \over \tau g^\prime{}^{\prime} (z)}}\,, 
\eqno(3.1.13)$$
where $z$ is a point in the interval $[a,b]$ where $g^\prime (z)=0$ and $g^\prime{}^{\prime} (x)= {d^2 g(x) \over dx^2}$. 
Setting the derivative of $g$ to zero we find that   
$$
\eta_{i} = - \xi_i ,\quad {\rm  and}\quad \hat \varphi = \zeta \quad {\rm or \ equivalently}\ \quad \hat \varphi_i=- \zeta_i
\eqno(3.1.14)$$
Putting everything together, in the saddle point approximation, we find that 
$$
{\Psi  (\zeta, \xi^i, a)  \equiv \lim_{\tau \to \infty} \tau^{{D-1\over 2}} \psi_\sigma (X) = - \lim_{\tau \to \infty} \left( {i\over 2\pi}  \right)^{{D-1\over 2}} \times  { m^{D-2}\over 2  \zeta^{D-1}} \times e^{i m \tau } \psi  (\zeta, \xi^i , a) }
\eqno(3.1.15)$$ 
\par
The massive  irreducible representation of the Poincar\'e group in momentum space is carried  by $\psi ({\bf p},a) $ in equation (3.1.3) which depends on ${\bf p}_i$, or equivalently $\eta_i$ and $\hat \varphi$, that is,  $\psi  (\hat \varphi , \eta_i , a) $.  
The wave function in position space depends on $X^\mu$, namely $\Psi (X ,a)$ but at time-like infinity which corresponds to taking $\tau\to \infty$ it depends on  hypersphere with parameters $\xi_i$ and $\eta$. Equation (3.1.14) tells us that the coordinates in momentum space and in position space are, up to a sign, equal.  Furthermore equation (3.1.15) tells us that  the wave function in momentum space is essentially equal to the wave function in position  space when we take $\tau\to \infty$. 
\par
Hence the massive  irreducible representations of the Poincar\'e group are completely described by functions on the hypersphere at 
$\tau\to \infty$ which carry a representation of ${\cal H}_{m\not =0}^{Poincare}$. As such at  the kinematic  level,  massive particles can be completely described by the behaviour of wave functions at $i^+$, the final hyperboloid in the foliation and so 
 in terms of local operators on this surface. 
\medskip
{\bf 3.2 Massive  $BMS_3$ irreducible representations  at time-like infinity}
\medskip
We now wish to generalise the above construction and push the massive irreducible  $BMS_3$ representations, found in section (3.1),  to time-like infinity.  As we discussed before,  massive in this context means they are the   analogous  representations  to the massive  irreducible representations  of the Poincar\'e group. As a first step let us review   the calculation given in previous section, but in three dimensions and using the notation of   the $BMS_3$ formulation. The general state of equation (3.1.3) can be written as 
$$
 \psi  ({\bf p},a ) = e^{\sum_j \hat \varphi_j J_{j0}} \psi  ({\bf p}^{(0)}, a)= e^{\varphi_{-1} J_1+  \varphi_{+1}J_{-1}} \psi  ({\bf p}^{(0)} , a)
\eqno(3.2.1)$$ 
Looking at equation (2.1.5), or the above equation,  and using equation (2.0.2) we identify 
$$
\hat \varphi_1= -(\varphi_{+1}+\varphi_{-1}), \ \hat \varphi_2= i(\varphi_{+1}-\varphi_{-1})\quad {\rm or \ equivalently}\quad
\varphi_{\pm 1}= -{1\over 2}(\hat \varphi_1\pm i\hat \varphi_2)
\eqno(3.2.2)$$
Recalling equation (2.1.8), namely  $x_0= X_0 , \quad x_{\pm 1}\equiv {1\over {2}}(X^2 \mp iX^1)$, we can write equations (3.1.8) and (3.1.10) as 
$$
x_0=  \tau y_0=\tau  \cosh \zeta ,  \quad x_{\pm 1}= \tau  y_{\pm 1}= \tau\zeta _{\pm 1} {\sinh \zeta\over \zeta}
\eqno(3.2.3)$$ 
where $ \zeta _{\pm 1} = {1\over 2}(\zeta^2\mp i\zeta^1)$ and $\zeta= 2\sqrt {\zeta_{+1}\zeta_{-1}}=\sqrt {(\zeta^1)^2+(\zeta^2)^2}$.
While,  using equation (3.2.2), equation (3.1.4)  can be written as 
$$
{\bf p}_0=m\cosh  \varphi ,\quad p_{\pm1}= {\bf p}_2\mp i {\bf p}_1= \pm 2i m\varphi_{\pm 1}{ \sinh  \varphi\over  \varphi}
 \eqno(3.2.4)$$
where $ \varphi\equiv \hat  \varphi=\sqrt {\hat \varphi_1^2+ \hat \varphi_2^2}=2\sqrt {\varphi_{+1}\varphi_{-1}}$. 
The reader can verify that for small $\varphi$ this agrees with equation (2.1.6). 
\par
In terms of these variables 
$$
X^\mu {\bf p}_\mu= \tau m g=m \tau ( \cosh \varphi \cosh \zeta + 2i {\sinh \varphi \over \varphi}{\sinh \zeta \over \zeta}(-\zeta_{1} \varphi_{-1}+ \zeta_{-1}\varphi_{+1}))
 \eqno(3.2.5)$$
  It is straightforward to take the derivatives of $g$,  as we did above, and find  that the contribution to the wave function in position space   as $\tau \to \infty$ is given when $\zeta_{\pm 1}= \mp 2i \varphi_{\pm 1}$. 
 \par
Let us now discuss how to push the irreducible representation of $BMS_3$ to time-like infinity. In section (2.1), we introduced a coordinate $x_n$ for each of the super momenta $p_n$ and the corresponding wavefunction was given in  equation (2.1.11). 
Examining equation (3.2.3) the above discussion of the Poincar\'e case can be naturally generalised to $BMS_3$ by taking 
 $$
 x_n= \tau y_n , \quad y^0= \cosh \zeta , \ y_n= \zeta _n {\sinh \zeta \over \zeta} ,\ n=\pm 1, \pm 2, \ldots 
 \eqno(3.2.6)$$
 where $\zeta=2 \sqrt{\sum_{n=1}^\infty \zeta_n \zeta_{-n}}$. As  $(x_n)^*= x_{-n} $ we find that    $(\zeta_n)^*= \zeta_{-n} $ for all $n$, although  $\zeta_0=0$. By abuse of notation we use the same symbol for $\zeta$. 
\par
 The expression for the wavefunction in position space  given in  equation (2.1.11) contains the quantity $e^{i \sum_n x_{-n}p_n}$ which is crucial for the behaviour at infinity and which we will now study.  Using equation (3.2.6) we can write
 $$
 \tau g(x_n ,p_n)\equiv  \sum_n x_{-n} p_n= x_0 p_0 +\sum_{n,  n\not=0} x_{-n}p_n = \tau (p_0\cosh \zeta  + {\sinh \zeta \over \zeta}\sum_{n,  n\not=0} \zeta_{-n}p_n )
  \eqno(3.2.7)$$
 \par
 The next step in pushing the Poincar\'e wavefunction to time-like infinity was to use the parameterisation of ${\bf p}_\mu$ of equation 
 (3.1.4), or equivalently (3.2.4). To do the same for $BMS_3$ we would need the analogue of these latter equations. Calculating the super momenta using equation (2.0.4) would give a parametrisation of $p_n$ in terms of $\varphi_n$ ($\varphi_0=0$) which would also solve equation (2.1.9). The lowest order solution was given in equation  (2.1.4) but the all orders solution is not known. 
 \par
 As such we can write $p_n (\varphi_n)$ and $x_n (\zeta_n)$ and then equation (2.1.11) takes the form 
 $$
\Psi (x_n,a)= \int \prod _n d\varphi_n J (\varphi_m) e^{i\tau g(\zeta_n, \varphi_n)} \psi (p_n (\varphi_m),a ) 
\eqno(3.2.8)$$
where $g(\zeta_n, \varphi_n)= g( x_n(\zeta_n),  p_{n}(\varphi_m))$. The contribution as $\tau \to\infty$ is then given by taking 
 $$
 {\partial g(\zeta_n, \varphi_n)\over \partial\varphi_n}=0 
  \eqno(3.2.9)$$
 \par
 We will now explicitly carry out this  calculation to lowest order in $\varphi_n$, which at  lowest order is related to the super momenta by  ${p_n\over m}= \delta _{n,0}+2in\varphi_n$,  and the coordinates $\zeta_n$. 
 In this case, setting $m=1$, we find that 
 $$
 g(\zeta_n, p_n)= p_0 (1+{\zeta^2\over 2}) + \sum_{n, n\not=0} \zeta_{-n} p_n
 = 1+  {1\over 2} \sum_{n, n\not=0} (\zeta_n+p_n)( \zeta_{-n} +p_{-n})
 \eqno(3.2.10)$$
 where we have used the fact, noted below equation (2.1.10),  that at lowest order 
 $$
 p_0=\sqrt {1+  \sum_{n, n\not=0} p_{-n} p_n}+\ldots =1+  {1\over 2} \sum_{n, n\not=0} p_{-n} p_n +\ldots 
  \eqno(3.2.11)$$
 The contribution as $\tau\to \infty$ is given by $\zeta_n= - p_n = -2in \varphi_n$ which agrees when restricted to the Poincar\'e case with the result of equation (3.1.14).  
 \par
Thus at lowest order we find that the coordinates at infinity, $\zeta_n$,  are essentially the same as the parameters,   $\varphi_n$,  which describe the momenta. It would seem inevitable that the solution of equation (3.2.9) will result in a one to one map between the spaces parameterised by $\varphi_n$ and $\zeta_n$. As such the massive  $BMS_3$ irreducible representation in position space  is described by a wavefunction that depends on the infinite number of coordinates $\zeta_n$ at time-like infinity. 
\par
In order to get a  better understanding of what is going on we introduced, in section (2.1),  a parameter $z$ in equation (2.1.13). In terms of the above variables we introduce 
 $$
  X(z)= \sum_m x_m z^{m} ,\ \ 
\ \ \zeta (z) = \sum _{n, n\not=0} \zeta_n z^n , \quad p (z) =  \sum _{n, n\not= 0} p_n z^n ,\quad \varphi (z) =  \sum _{n, n\not=0} \varphi_n z^n
  \eqno(3.2.12)$$
  We note that  $(X(z))^*= X(z)$,  $\zeta (z)^*=  \zeta (z)$ and similarly for the other quantities.  
  \par
The lowest order relation between the coordinates $\zeta_n $ at time-like infinity and the $\varphi_n$  which parameterise the super momenta was given below equation (3.2.11) and this can be written as 
 $$
 \zeta (z)= - p(z)  = -2iz{d\over dz}  \varphi (z)
\eqno(3.2.13)$$
although the relation at higher orders may be  more complicated. 
\par
In terms of these variables the wave function in x  space of the massive $BMS_3$ irreducible representation at time-like infinity  depends on $\zeta(z)$, that is, $\Psi (\zeta(z), a)$. To  clarify the meaning of this dependence it is useful to introduce the variables $X^1_n$, $X^2_n$, or equivalently 
 $\zeta ^1_n$, $\zeta ^2_n$, as well as $p_{1n}$ , $p_{2n}$  for $n=1,2,\ldots $. These are related to the above variables by 
$$
X_{\pm n}= {1\over 2} (X^2_n\mp i X^1_n) , \quad p_{\pm n}= p_{2n}\mp i p_{1n} , \quad \zeta_{\pm n}= {1\over 2} (\zeta^2_n\mp i \zeta^1_n)
\eqno(3.2.14)$$
We identify the coordinates of the original spacetime as $X^1= X^1_1$,  $X^2= X^2_1$ or $\zeta ^1= \zeta^1_1$ and $\zeta ^2= \zeta^2_1$. In terms of these variables we have 
$$
\zeta (z)= \sum_{n=1}^\infty (\zeta^2_n \cos n\theta + \zeta^1_n \sin n\theta ) \equiv \zeta^2 (z) + \zeta^1 (z)
= \zeta^2 \cos \theta + \zeta^1 \sin \theta +\ldots 
\eqno(3.2.15)$$
where $z=e^{i\theta}$. We can think of $z$, or $\theta$,  as a parameter of a closed string  As we take $\theta$ from $0$ to $2\pi$ we move in spacetime in a circle in the $\zeta^1$, $\zeta^2$ original space with higher modes in the higher coordinate $\zeta ^1_n$, $\zeta ^2_n$ space. 
\par
We can interpret  the massive extended extended $BMS_3 $ irreducible representation at time-like infinity as a string 
living in the hyperboloid at time-like infinity which has coordinates $\zeta^1$ and $\zeta^2$  with $\theta$ being the parameter on the string. The higher order coordinates $\zeta^1_n$ and $\zeta^2_n$ describe the higher modes of the string. Equation (2.1.4) tells us that the super rotations act so as to change the parameterisation of the string. Thus the massive  $BMS_3$ irreducible representation is described at time-like infinity by a one dimensional extended object, a string, living in the two dimensional hyperboloid  at time-like infinity.


\medskip
{\bf 4. Extended $BMS_4$ irreducible  representations }
\medskip
In this section we will find  irreducible representations of  extended $BMS_4$ which has the super rotation generators $J_n$, $\bar J_n$, $n\in Z$, and the supertranslations generators $P_{r,s}$ and $\bar P_{r,s}$, $r, s\in Z+{1\over 2}$. Their  algebra was written in [5]  and we use the conventions of [24,38]. 
$$
[ J_n, J_m]= (n-m) J_{n+m} , \quad [ \bar J_n, \bar J_m]= (n-m) \bar  J_{n+m} , \quad [ J_n, \bar J_m]=0
\eqno(4.0.1)$$
and 
$$
[ J_n, P_{k,l} ]= ({n\over 2}-k ) P_{k+n,l} ,\quad [ \bar J_n, P_{k,l} ]= ({n\over 2}-l ) P_{k,l+n} , \quad [P_{k,l} , P_{r,s}]=0
\eqno(4.0.2)$$
\par
The Poincar\'e subalgebra ${\bf P}_\mu, {\bf J}_{\mu\nu}$, $\mu,\nu=0,1,2,3,4$  is generated by 
$$
{\bf J}_{13}= -{1\over 2} (J_{1}+J_{-1} +\bar J_{1}+\bar J_{-1} ) , \ {\bf J}_{23}= {i\over 2} (-J_{1}+J_{-1} +\bar J_{1}-\bar J_{-1} ) . \ 
{\bf J}_{12}= i(J_0-\bar J_0)
$$ 
$$
{\bf J}_{01}= {1\over 2} (-J_{1}+J_{-1} -\bar J_{1}+\bar J_{-1} ) , \ {\bf J}_{02}= -{i\over 2} (J_{1}+J_{-1} -\bar J_{1}-\bar J_{-1} ) . \ 
{\bf J}_{03}= J_0+\bar J_0
\eqno(4.0.3)$$
and 
$$
{\bf P}_0= P_{{1\over 2}, {1\over 2}}+ P_{-{1\over 2}, -{1\over 2}} , \ {\bf P}_1=-( P_{-{1\over 2}, {1\over 2}}+ P_{{1\over 2}, -{1\over 2}}) , \ 
$$
$$
{\bf P}_2=-i( P_{{1\over 2}, -{1\over 2}}- P_{-{1\over 2}, {1\over 2}} ) ,\ {\bf P}_3= -(P_{{1\over 2}, {1\over 2}}- P_{-{1\over 2}, -{1\over 2}} )
\eqno(4.0.4)$$
\par
The momentum eigenvalues $p_{k,l}$ transform as 
$$
\delta p_{k,l}= - \sum_n ({n\over 2} -k ) \Lambda _{-n} p_{n+k,l}  - \sum_n ({n\over 2} -l ) \bar \Lambda _{-n} p_{k,n+l} 
\eqno(4.0.5)$$
under the super rotation  $e^{\sum_m(\Lambda_{-m} J_m +\bar \Lambda_{-m} \bar J_m) }$. 
\medskip
{\bf 4.1 Massive extended $BMS_4$ representations }
\medskip
We will now construct the massive extended $BMS_4$ irreducible representations  meaning the ones that correspond to the massive irreducible representation of the Poincar\'e group. In the latter case we can choose to take the rest frame momentum to be  ${\bf p}^{(0)}_0=m$, with all other components zero. In the extended $BMS_4$ case we take only the corresponding super momentum  to be non-zero which is  
$$
p^{(0)}_{{1\over 2}, {1\over 2}}={m\over 2} = p^{(0)}_{-{1\over 2}, -{1\over 2}}
\eqno(4.1.1)$$
Taking the super rotations to be of  the form $e^{\sum_m (\Lambda _{-m} J_m+ \bar \Lambda_ {-m} \bar J_m)}$ we find, using equation (4.0.5),  that this choice is preserved if 
$$
\delta p_{k,l}=-{m\over 2} ( \delta_{l, {1\over 2}} {(1-6k)\over 4} \Lambda _{k-{1\over 2} }
-\delta_{l,- {1\over 2}} {(1+6k)\over 4} \Lambda _{k+{1\over 2} }
$$
$$
+\delta_{k, {1\over 2}} {(1-6l)\over 4} \bar \Lambda _{l-{1\over 2} }
-\delta_{k,- {1\over 2}} {(1+6l)\over 4}\bar  \Lambda _{l+{1\over 2}} )=0
\eqno(4.1.2)$$
It is straightforward to show that this implies that $\Lambda _p $ and $\bar \Lambda _p $  vanish for all $p$ except for $p=0$ whose
parameters  obey the condition 
$$
\Lambda_0+\bar \Lambda_0=0
\eqno(4.1.3)$$
\par
Thus the isotropy group contains the one generator ${\cal H}_{m\not=0} ^4= i(J_0-\bar J_0)= J_{12}$. The irreducible representation is built out of a single state which is an eigenstate of $J_{12}$. The most general state in the irreducible representation is given by boosting this state with 
$e^{\sum_m(\varphi_{-m} J_m +\bar \varphi_{-m} J_m )}$ with $\varphi_0=\bar \varphi_0$
\par
This is quite different from what one might expect if one was thinking about the massive irreducible representation of the Poincar\'e subgroup. To understand more clearly what is going on we will repeat the above calculation but for just the Poincar\'e group. As such we take  only the  parameters 
$\Lambda_0$ , $\bar \Lambda_0$,  $\Lambda_{\pm 1}$, $\bar \Lambda_{\pm 1}$ to be non-zero as well as  considering only the momenta
$p_{\pm {1\over 2}, \pm {1\over 2}}$ and $p_{\pm {1\over 2}, \mp {1\over 2}}$. Putting these restrictions into equation (4.1.2) we find the three conditions 
$$
\Lambda_0+\bar \Lambda_0=0 , \ \Lambda_{-1}-\bar \Lambda_{1}=0 ,\  \Lambda_{1}-\bar \Lambda_{-1}=0 
\eqno(4.1.4)$$
Thus the isotropy group  has the generators $J_0-\bar J_0$, $J_{-1}+\bar J_{1}$ and $J_{1}+\bar J_{-1}$ which we can also write  as  
$$
J_{12}=i(J_0-\bar J_0) ,\ J_{13} = -{1\over 2}(J_{1}+\bar J_{-1} +J_{-1}+\bar J_{1})  , \    J_{23}= -{i\over 2}(J_{1}+\bar J_{-1} -J_{-1}-\bar J_{1})
\eqno(4.1.5)$$
These   generate SO(3) as indeed had to be the case as this is the isotropy group for the massive particle of the Poincar\'e group. 
\par
The irreducible representation of the massive Poincar\'e group is constructed from a state $|p^{(0)}, a\rangle $ which carries a representation of SO(3). The general state is boosted by the generators $J_{01}$, $J_{02}$ and $J_{03}$  which, in terms of the above notation,   can be written as 
$$
|p,a\rangle = g_1 |p^{(0)}, a\rangle \ {\rm where }\ g_1 (\chi)=e^{\chi (J_1- \bar J_{-1} ) + \bar \chi ( J_{-1}-\bar J_{1} )
+ \chi_0 (J_0+\bar J_0)}
\eqno(4.1.6)$$
\par
We get a smaller isotropy group for extended $BMS_4$ as even the usual Lorentz rotations act on  some of the higher super momenta to give   additional constraints, for example 
$$
\delta P_{-{3\over 2} , -{1\over 2}}= m\Lambda_{-1}  =0
\eqno(4.1.7)$$
Thus  the massive   irreducible representation of the Poincar\'e group  is constructed from a representation of SO(3) while the 
 the massive  irreducible representations of extended $BMS_4$ is constructed out of a representation of the much smaller SO(2). As such  these latter  irreducible representations do not contain the massive irreducible representations of the Poincar\'e group! 
\par

The general state in the irreducible massive representation of extended $BMS_4$ is given by
$$
| p_{r,s}, a\rangle =g_2 (\varphi) g_1 (\chi) g_3(\phi) | p^{(0)}, \lambda \rangle \quad {\rm where }\quad  (J_0 -\bar J_0)| p^{(0)}, \lambda \rangle= \lambda | p^{(0)}, \lambda \rangle
\eqno(4.1.8)$$
with  
$$
g_2(\varphi)=e^{\sum_{m, |m|\ge 2} (\varphi_{-m}J_{m} +\bar \varphi_{-m}\bar J_{m} )}, \quad 
g_3 (\phi )= e^{\phi (J_1+ \bar J_{-1} ) + \bar \phi ( J_{-1}+\bar J_{1} )}
\eqno(4.1.9)$$
We recognise $g_1(\chi)$ as the boost for the Poincar\'e group, $g_3(\phi)$ as the boost arising because  the isotropy group for extended $BMS_4$ is smaller and $g_2$ as the boost involving generators not in the Poincar\'e group. The group element $g_2g_1g_3$
belongs to the coset of extended $BMS_4$ with subgroup $J_0-\bar J_0$. 
\par
Examining equation (4.1.8) we see that the general extended $BMS_4$ state depends on $\varphi_m$ and $\bar \varphi_m$, $|m|\ge 2$, and $\chi , \bar \chi , \chi_0$,  which correspond to  the same boosts as for the Poincar\'e group, it  also depends on $\phi, \bar \phi$ which correspond to Poincar\'e generators which were in the SO(3) isotropy group of  the massive Poincar\'e irreducible representation.  As such the general  extended $BMS_4$ state has  an additional dependence, that is on  $\phi$ and $\bar \phi$  to that  found in the case of the Poincar\'e group in addition to the $\varphi$ and $\bar \varphi$ dependence  arising from the generators not in the Poincar\'e group.  
\par
Although  the massive  irreducible representations of  extended $BMS_4$  does not contain the massive  irreducible representations  of the Poincar\'e group, we can construct   reducible representations of extended $BMS_4$ that do contain the irreducible representations  of the Poincar\'e group. We take a representation of the SO(3)  group,  which was  the isotropy group of the Poincar\'e group,  to be  carried by the states $ | p^{(0)}_{r,s}, a\rangle$ and boost it by the extended $BMS_4$ transformation to find the general state  as follows 
$$
| p_{r,s}, a\rangle =g_2 (\varphi)g_1(\chi) | p^{(0)}_{r,s}, a\rangle
\eqno(4.1.10)$$
We note that this boost is now missing the Poincar\'e group element $g_3(\phi)$ which is now  in the new SO(3) "isotropy group". The group element $g_2(\varphi)g_1(\chi)$ belongs to the coset of extended $BMS_4$ with subgroup SO(3). 
\par
Under a super rotation the general state transforms as 
$$
e^{\sum_m (\Lambda _{-m} J_m+ \bar \Lambda_ {-m} \bar J_m)} | p_{r,s}, a\rangle
= g_2(\varphi^\prime) g_1(\chi^\prime) h (\varphi, \chi)  |  p^{(0)}_{r,s}, a\rangle
\eqno(4.1.11)$$
where $h (\varphi, \chi) $ is an element of  SO(3) which acting on the rest frame states leads to  a matrix transformation. 
\par
Let us now compute the momentum of the general state of the irreducible representation which we can rewrite in the generic form as 
$$
|p_{k,l}\rangle = e^{\sum_n \varphi_{-n} J_n} e^{\sum_n \bar \varphi_{-n} J_n} |p_{k,l}^{(0)}\rangle
\eqno(4.1.12)$$
The $\varphi_n$ and  $\bar \varphi_n$ are subject to certain  constraints, which for the massive case is $\varphi_0 =\bar \varphi_0$,  but we will do the calculation without initially specifying the constraints so that it applies to the general case. 
The momenta are then given by 
$$
p_{k,l}= p^{(0)}_{k,l}-\sum_n \varphi_{-n} ({n\over 2}-k) p^{(0)}_{k+n,l}
$$
$$
+{1\over 2} \sum_{n,m} \varphi_{-n}\varphi_{-m} ({n\over 2}-k) ({m\over 2}-k-n) p^{(0)}_{k+n+m,l}+\ldots + (\varphi_n \to \bar \varphi_n , k\leftrightarrow l )
\eqno(4.1.13)$$
Thus we find an expression for the super momenta $p_{k,l}$ in terms of $\varphi_n$ and  $\bar \varphi_n$, but there are many more of the former than the latter and so there must be an infinite number of constraints on the super momenta. 
\par
For the massive case the rest frame momentum is given by equation (4.1.1) and substituting  this into the above equation we immediately find that  $p_{k,l}=0 $ unless either $k$ or $l$ takes the values $\pm {1\over 2}$. One finds at  lowest order that the non-zero super momenta are given by 
$$
p_{{1\over 2}, {1\over 2}}= {m\over 2}(1+{1\over 2}(\varphi_0+\bar \varphi_0)) ,\quad p_{{1\over 2},- {1\over 2}}={m\over 2}( \varphi_1-\bar \varphi_{-1}) ,\quad p_{{1\over 2}, l}= -{m\over 2}{(1-6l)\over 4} \bar\varphi_{-{1\over 2} +l} , l\not= \pm {1\over 2} 
$$
$$
p_{-{1\over 2}, -{1\over 2}}= {m\over 2}(1-{1\over 2}(\varphi_0+\bar \varphi_0)) ,\quad p_{-{1\over 2}, {1\over 2}}={m\over 2}(- \varphi_{-1}+\bar \varphi_{1}) ,\quad
$$
$$
 p_{-{1\over 2}, l}= {m\over 2}{(1+6l)\over 4} \bar\varphi_{{1\over 2} +l} , l\not= \pm {1\over 2} 
\eqno(4.1.14)$$
All the super momenta of the form $p_{\pm{1\over 2}, l}$ are determined by  $\bar \varphi_n$. We observe that up to this order $p_{{1\over 2}, l+1}={(5+6l)\over (1+6l)} p_{-{1\over 2}, l}$.   The equivalent results for $p_{\pm{1\over 2}, l}$ follow from the above if we swap $k$ and $l$, $\varphi _n\leftrightarrow \bar \varphi_n$ and take $p_{k,l}\leftrightarrow p_{l,k}$ and we conclude that 
$p_{k, \pm{1\over 2}}$ are determined by $\varphi_n$.  It is clear that the super momenta must satisfy an infinite number of constraints whose form at lowest order could be computed from the above expressions when extended to higher orders.

\medskip
{\bf 4.2 Massless  extended $BMS_4$ representations }
\medskip
We now construct the massless irreducible representations of extended $BMS_4$  and so choose  our rest frame momentum to be 
$$
{\bf p}^{(0)}_{-}\equiv {1\over \sqrt {2}}(p^{(0)}_3-p^{(0)}_0)= 1 , \ {\rm or \ equivalently}\ p^{(0)}_{{1\over 2}, {1\over 2}}=- {1\over \sqrt {2}}
\eqno(4.2.1)$$
all other super momenta being zero. 
Using equation (4.0.5) we find this is preserved provided 
$$
\delta p_{k,l}={1\over \sqrt {2}}(\delta_{l,{1\over 2}}\Lambda_{k-{1\over 2}} {(1-6k)\over 4}
+\delta_{k,{1\over 2}}\bar \Lambda_{l-{1\over 2}} {(1-6l)\over 4})
=0
\eqno(4.2.2)$$
From this  we find that $\Lambda_p=0$ and  $\bar \Lambda_p=0$ for $p\not=0$ as well as  the condition 
$$
\Lambda_0+\bar \Lambda_0=0
\eqno(4.2.3)$$
Thus the isotropy group ${\cal H}^4_{m=0}$ contains one generator, namely $J_0-\bar J_0$. This is the same isotropy group as for the massive case. 
\par
The irreducible representation is built out of  a single state $| p^{(0)}_{k,l}, \lambda\rangle $ which is an eigenstate of this generator, that is, $(J_0-\bar J_0)| p^{(0)}_{k,l}, \lambda\rangle=\lambda | p^{(0)}_{k,l}, \lambda\rangle$. The general state is found by boosting
with the generators 
$ J_m$, $\bar J_m $ $|m|\ge 1$ and  $J_0 +\bar J_0$. We can write this as 
$$
  | p_{k,l}, \lambda\rangle = 
  e^{\sum_{m, |m| \ge 2}  (\varphi_{-m} J_m +\bar \varphi_{-m} \bar J_m )} e^{\chi (J_{1}+ \bar J_{1})+\bar \chi (J_{1}-\bar J_{1})+\chi_0(J_0+\bar J_0)}
  $$
  $$
e^{\phi(J_{-1}+ \bar J_{-1})+\bar \phi(J_{-1}-\bar J_{-1}) }  | p^{(0)}_{k,l}, \lambda\rangle
\equiv g_2(\varphi)  g_1(\chi) g_3(\phi) | p^{(0)}_{k,l}, \lambda\rangle
\eqno(4.2.4)$$
\par
In order to better understand what is going on it is useful to analyse the situation for the Poincar\'e group in the same notation. In this case we only consider $p_{\pm {1\over 2}, \pm {1\over 2}}$ and $p_{\pm {1\over 2}, \mp {1\over 2}}$ and the generators $J_0$, $J_{\pm 1}$,  $\bar J_0$ and $\bar J_{\pm 1}$. As such we only take  the   parameters 
$\Lambda_0$, $\bar \Lambda_0$,  $\Lambda_\pm$, $\bar \Lambda_\pm$ to be non-zero.  
Equation (4.2.2)  then  implies that 
$$
\Lambda_0 +\bar  \Lambda_0=0 , \ \Lambda_{-1}=0= \bar \Lambda_{-1}
\eqno(4.2.5)$$
Hence the isotropy group ${\cal H}^{4dPoincare}_{m\not=0}= SO(2)\otimes_s T^2$ contains the generators $J_0-\bar J_0$, $J_{-1}$ and $\bar J_{-1}$ which we can write as 
$$
J_{12}=i(J_0-\bar J_0), \  J_{-1}+\bar J_{-1} =- \sqrt {2} J_{1+} ,\   J_{-1}-\bar J_{-1} =- i\sqrt {2} J_{2+} 
\eqno(4.2.6)$$
leaving the generators 
$$
J_0+\bar J_0= J_{03}= J_{+-} , \  J_{1}+\bar J_{1} =- \sqrt {2} J_{1-} ,\   J_{1}-\bar J_{1} = i\sqrt {2} J_{2-} 
\eqno(4.2.7)$$
from which  the boosts will be constructed.
\par
The usual   irreducible representations of the Poincar\'e group which correspond to conventional  particles with discrete spin are  constructed from the state $|p^{(0)}, \lambda \rangle $ which obeys 
$$
(J_0-\bar J_0)|p^{(0)}, \lambda \rangle=\lambda |p^{(0)}, \lambda \rangle , \quad  J_{-1}|p^{(0)}, \lambda \rangle=0, \ \quad 
\bar J_{-1} |p^{(0)}, \lambda \rangle=0
\eqno(4.2.8)$$
Adopting the latter constraints means that the effective isotropy group is generated by $J_{12}=i(J_0-\bar J_0)$ and so is SO(2) which is the same as for the massless extended $BMS_4$ irreducible representation. 
The general state is given by 
$$
|p,\lambda \rangle= g_1(\chi) |p^{(0)}, \lambda \rangle
\eqno(4.2.9)$$
\par
Returning to the extended $BMS_4$ massless irreducible representation we observe that if we were to take the rest state to obey 
$$
J_{-1}|p^{(0)}_{r,s}, \lambda \rangle=0, \ 
\bar J_{-1} |p^{(0)}_{r,s}, \lambda \rangle=0
 \eqno(4.2.10)$$
then the general state of equation (4.2.4) takes the form of $ | p_{r,s}, \lambda\rangle= g_2(\varphi) g_1(\chi) |p^{(0)}_{r,s}, \lambda \rangle$. Since $g_1$ is the boost of the Poincar\'e massless particle we can think of this as a Poincar\'e state which is boosted by the higher mode generators of  extended $BMS_4$ contained in $g_2(\varphi)$. As such the massless extended $BMS_4$ irreducible representations do contain the   massless irreducible representations of the Poincar\'e group. 
\par
There are other irreducible representations of the Poincar\'e group called the continuous spin representations. These do not satisfy the last two conditions in equation (4.2.8). 
\par
Given the general state can be written in the form of equation (4.1.12),  subject to the constraint $\varphi_0=\bar \varphi_0$ , and  that the rest frame momenta given in  equation (4.2.1), we can compute its super momenta to find that 
$$
p_{k,l}=-{1\over \sqrt 2} (\delta_{k,{1\over 2}}\delta_{l,{1\over 2}} - \delta_{l,{1\over 2}}  {(1-6k)\over 4}\varphi_{-{1\over 2}+k}
- \delta_{k,{1\over 2}}  {(1-6l)\over 4}\bar \varphi_{-{1\over 2}+l}+ \ldots )
\eqno(4.2.11)$$
Hence $p_{k,l}$ is zero unless either $k$ or $l$ is equal to ${1\over 2}$. The non-zero components at lowest order are 
$$
p_{{1\over 2}, l}=-{1\over \sqrt 2} ( \delta_{l,{1\over 2}}(1+{1\over 2} \varphi_0)-{(1-6l)\over 4} \bar \varphi_{-{1\over 2}+l}), \quad 
$$
$$
p_{k, {1\over 2}}= -{1\over \sqrt 2} (\delta_{k,{1\over 2}}(1+{1\over 2} \bar \varphi_0)-{(1-6k)\over 4}  \varphi_{-{1\over 2}+k})
\eqno(4.2.12)$$
As such $p_{{1\over 2}, l}$ and $p_{k, {1\over 2}}$ are determined by $\bar \varphi_n$ and $\varphi_n$ respectively. 
\par
It would be interesting to find the isotropy group for other momenta that are not in the orbit of the massive and massless cases. 
An interesting case is to take $p_{-{3\over 2}, -{3\over 2}}=1$. Looking at equation (4.0.5) we find three group elements with non-zero $\Lambda$'s, $\bar \Lambda$' s given by  $\Lambda_{1}=1$, $ \bar \Lambda_{1}=1$ and $\Lambda_0=-\bar \Lambda_0=1$, that is,  the algebra SO(3). The corresponding parameters can also be written as $\Lambda (z)= z$   $\bar \Lambda (z)= 0$ , $\Lambda (z)=0$, $\bar \Lambda (z)= \bar z$,  and 
 $\Lambda (z)= 1=-\bar \Lambda (\bar z)$  and the reader can verify that they provide three  solutions for the isotropy algebra by using equation (4.3.4). We note that this is not one of the momenta in the orbit of the massive case, see equation (4.3.9).\footnote {{$\dagger$}}{We thank Blagoje Oblak for drawing our attention to this case.}

\medskip 
{\bf 4.3 Interpretation of the extended $BMS_4$  representations}
\medskip
The above results are summarised at the beginning of section six. 
At the end of  section  (2.1) and in section (3.2) we argued that the massive  irreducible representations of $BMS_3$ were not carried by a particle but by an extended one dimensional object. In this section we give the analogous  discussion  for  the massive  irreducible representations of extended $BMS_4$. Given the wave function in super momentum space $\psi (p_{r,s}) $ we can  find the wave function in x space. To do this we must introduce  x space coordinates $x_{r,s}$ for  each of the super momenta $p_{r,s}$. 
The wave function  in x space has the generic form 
 $$
\int d\varphi J (\varphi) e^{i \sum _{r,s} p_{r,s}x_{-r,-s}}\psi (p_{r,s})
\eqno(4.3.1)$$
The coordinates $x_{r,s}$  transform as 
$$
\sum_{r,s} x_{-r,-s}^\prime P_{r,s} = e^{-\sum_m  \Lambda_{-m} J_m+ \bar \Lambda_{-m} \bar J_m} \sum_n x_{-r,-s} P_{r,s} e^{\sum_m \Lambda_{-m} J_m+ \bar \Lambda_{-m} \bar J_m}  \Rightarrow 
$$
$$
\delta x_{r,s} =-\sum_m ({3\over 2}m +r) \Lambda_{-m} x_{r+m,s}-\sum_m ({3\over 2}m +s) \bar \Lambda_{-m} x_{r,s+m}
\eqno(2.3.2)$$
\par
Like in the three dimensional case we introduce parameters $z=e^{i\theta} $ and $\bar z=e^{i\bar \theta}$ and define 
$$
p(z,\bar z) = \sum_{r,s } p_{r,s} z^{r}\bar z^{s} ,\quad \Lambda (z) = \sum_{m} \Lambda_m z^{m}  ,  \quad 
 \bar \Lambda (\bar z) = \sum_{m}\bar  \Lambda_m \bar z^{m} ,\ x(z,\bar z)= \sum_{r,s} x_{r,s} z^r \bar z^s
\eqno(4.3.3)$$
The transformation of the super momenta of equation (4.0.5) can be written as 
$$
\delta p(z,\bar z)= {3\over 2} z{d\Lambda (z)  \over dz} p(z,\bar z)+  \Lambda (z) z{d p(z,\bar z)  \over dz} 
+{3\over 2} \bar z{d\bar \Lambda (\bar z)  \over d\bar z} p(z,\bar z)  +\bar \Lambda (\bar z) \bar z {d p(z,\bar z)  \over d\bar z} 
\eqno(4.3.4)$$
While the transformation of the coordinates is given by 
$$
\delta x(z,\bar z) = {1\over 2} z{d\Lambda (z) \over dz} x(z,\bar z) -z{d x(z,\bar z)  \over dz}\Lambda (z) 
+{1\over 2} \bar z{d\bar \Lambda (\bar z) \over d\bar z} x(z,\bar z) -\bar z{d x(z,\bar z)  \over d\bar z}\bar \Lambda (\bar z) 
\eqno(4.3.5)$$
Thus the super rotations of extended $BMS_4$ are  a reparameterisation $x(z,\bar z)$ under the parameters of $z$ and $\bar z$ but separately in $z$ and $\bar z$.  
\par
In section 3.1 we argued that the massive irreducible representation of extended $BMS_3$ was described by a string and we will give the analogous discussion here for extended $BMS_4$. What the above very generic discussion does not take account of is that one must introduce a position coordinate for every independent momenta, or equivalently every $\varphi_n$ and $\bar \varphi_n$. For both the  above massive and massless representations the momenta are very highly constrained as is apparent from equations (4.1.14) and (4.2.12) respectively and as such we must only introduce the corresponding position space   coordinates. In particular for the massive case the momenta $p_{r,s}$ are only non-zero if either $r$ or $s$ are equal to $\pm {1\over 2}$. As such, and to make contact with our usual spacetime, we introduce the quantities 
$$
p_{{1\over 2}, r}= {1\over 2}( p_{0, r}-p_{3 , r}), \ p_{-{1\over 2}, -r}= {1\over 2}( p_{0, r}+p_{3 , r}),\ 
$$
$$
p_{r, -{1\over 2}}= {1\over 2}( -p_{1,r}+ip_{2, r}),\ p_{-r, {1\over 2}}=- {1\over 2}( -p_{1,r}+ip_{2 ,r}) , \quad r\ge {1\over 2}
\eqno(4.3.6)$$
where we recognise the momenta of our usual spacetime as 
$$
p_{0 ,{1\over 2}} =p_0,\  p_{3 ,-{1\over 2}} =p_3 ,\ p_{{1}, {1\over 2} }  =-p_1, \  p_{{2}, -{1\over 2} }  =p_2
\eqno(4.3.7)$$
We also introduce 
$$
x_{-{1\over 2},- r}= ( x^0_{r}-X^3_{ r}), \ x_{{1\over 2}, r}= ( x^0_{r}+x^3_{ r}),\ 
$$
$$
x_{-r, {1\over 2}}= ( x^1_{r}-iX^2_{ r}),\ x_{r, -{1\over 2}}=( -x^1_{r}+ix^2_{ r}) , \quad r\ge {1\over 2}
\eqno(4.3.8)$$
\par
Then 
$$
\sum_{r,s} p_{r,s} x_{-r,-s}= \sum_{r\ge {1\over 2}} (p_{0,r} x^0_r+ p_{3,r} x^3 _r + p_{2,r} x^2 _r + p_{1,r} x^1 _r )
\eqno(4.3.9)$$
$$
x^0_{{1\over 2}}= X^0, \ x^3_{{1\over 2}}= X^3,\ x^1_{{1\over 2}}= X^1,\ x^2_{{1\over 2}}= X^2,
\eqno(4.3.10)$$
\par
The above quantities can be naturally encoded in two quantities if we introduce parameters $z$ and $\bar z$, namely 
$$
\bar p(\bar z)= \sum_{r\ge {1\over 2}}( p_{{1\over 2} , r} \bar z^r+ p_{-{1\over 2} ,- r} \bar z^r ) , \ 
p( z)= \sum_{r\ge {1\over 2}}( p_{r,-{1\over 2}} z^r+ p_{-r, {1\over 2} } z^r )
\eqno(4.3.11)$$
and 
$$
\bar X(\bar z)= \sum_{r\ge {1\over 2}}( x_{{1\over 2} , r} \bar z^r+ x_{-{1\over 2} ,- r} \bar z^r ), \ 
X( z)= \sum_{r\ge {1\over 2}}( x_{r,-{1\over 2}} z^r+ x_{-r, {1\over 2} } z^r )
\eqno(4.3.12)$$
Then 
$$
\sum_{r,s} x_{-r,-s} p_{r,s} = \int { dz \over z}p(z) X({1\over z}) +  \int { d\bar z \over\bar  z}\bar p(\bar z) \bar X({1\over \bar z}) 
\eqno(4.3.13)$$  
These expansions are guided by the solutions of the momenta in terms of $\varphi_n$ and $\bar \varphi_n$ given in equation (4.1.14). 
\par
Taking the Fourier transformation we find the massive irreducible of extended $BMS_4$ is a functional which depends on $X(z)$ and $\bar X(\bar z)$. As such this representation is carried by a string. 
\par 
It would be interesting to push this representation to time-like infinity as we did in  three dimensions. To do this we should express 
equation (4.3.13) as a square, like in equation (3.2.10) and then differentiate to find the relationship between the independent momenta and the coordinates on the hyperboloid  at time-like infinity.   One should find that the above  representation is carried by a string living on the three dimensional hyperboloid  at time-like infinity.


\medskip
{\bf 5 Irreducible representations of BMS without the super rotations}
\medskip
In this section we will construct the irreducible representations of BMS symmetries without including the super rotations. 
We begin with ${\bf BMS_3}$. In this case we have the Lorentz generators $J_0$ and $J_{\pm 1}$ and all the super translations $P_n$. As such only the parameters $\Lambda_0$ and $\Lambda_{\pm 1}$ are non-zero. Clearly the isotropy group can only be the same or smaller than that  for the BMS group. 
\par
For the {\bf massive} irreducible  representation,  considered for $BMS_3$ in section (2.1.1),  we find,  examining equation (2.1.1), that only $\Lambda_0$ is non-zero and so  the isotropy group just contains the generator  $J_0$. This is the same isotropy group  as we had when the super rotations were present and indeed we also had for the massive  Poincar\'e particle. 
\par
The  general state can be taken to be 
$$
|  p_n,a \rangle \equiv  e^{\varphi_{+1} J_{-1}}e^{\varphi_{-1} J_{1}} |  p_n^{(0)},a \rangle
\eqno(5.1.1)$$
One finds that the super momenta $p_n$ are given by 
$$
p_n = (n+1)m  (i\varphi_{+1} )^n(1-2\varphi_{+1}\varphi_{-1}) ,   n\ge 0 , \quad p_{-1}=-2i m\varphi_{-1} 
\eqno(5.1.2)$$
The result for the remaining super momenta is more complicated, one finds that 
$$
p_{-n}=m {(-\varphi_{+1}\varphi_{-1})^n\over (n-2)!} \sum _{p=0}^\infty {(n+p+2)!\over m!}(n+p+1) (-i\varphi_{-1})^p , \quad n \ge 2 
\eqno(5.1.3)$$
\par
Since we have an infinite number of super momenta but the general state only depends on the  two variables, $\varphi_{\pm 1}$, the super momenta must obey an infinite number of constraints that can be found from the above expressions for the momenta. Indeed we can express all the super momenta in terms of  $p_{\pm 1}$.  As a result  the irreducible representation is rather similar to that for the Poincar\'e group. 
\par
The {\bf massless $BMS_3$} irreducible representation was the subject of section (2.2) and  looking at equations (2.2.2) and (2.2.3) we conclude that $\Lambda_p= 0$ for all $p=0, \pm 1$. In particular we find that $\delta p_2= -{3im\over 2}\Lambda_ {1}=0$ and $\delta p_{-2}= -{3im\over 2}\Lambda_ {-1}=0$ and so $\Lambda_{\pm 1}=0$ which, with the variations of $p_{\pm 1}$,  lead to the 
 result that  there is no isotropy group in contrast to the case when the super rotations were present which has an isotropy group with two generators while  the massless Poincar\'e particle which has an isotropy group with the  one generator $J_{1+}$. However, in the usual representations this latter generator is taken to vanish on the rest frame state. 
We can write the general state for $BMS_3$ without super rotations  as 
$$
|p\rangle = e^{\varphi _{-}J_{1+}}e^{ \varphi_{ +}J_{1-} } e^{\phi J_{+-}} |p^{(0)}\rangle
\eqno(5.1.4)$$
where $J_{1-}= -{1\over \sqrt {2}} (J_0+{1\over 2}(J_1+J_{-1}))$,  $J_{1+}= {1\over \sqrt {2}} (-J_0+{1\over 2}(J_1+J_{-1}))$ and 
$J_{+-}=J_{02}=-{i\over 2} (J_1-J_{-1})$ which are Lorentz generators in light-cone notation. It will coincide with the Poincar\'e case if we also take $J_{1+}$ to vanish on the rest frame state. 
\par
We now consider the same calculations for extended ${\bf BMS_4}$,  but without the super rotations. We now have only the Lorentz generators $J_0$, $J_{\pm 1}$ and  $\bar J_0$, $\bar J_{\pm 1}$ and so only the corresponding parameters in the group transformations are now non-zero. The {\bf massive} irreducible representations of $BMS_4$ were given in section (4.1). Examining equation (4.1.2) we find that $\Lambda_{\pm 1}=0$
and  $\bar \Lambda_{\pm 1}=0$
leaving only $\Lambda_0$ and $\bar \Lambda_0$ which obey the condition $\Lambda_0+\bar\Lambda_0=0$. Thus the isotropy group is generated by $J_0-\bar J_0$. Thus we find the same isotropy group as when the super rotations were present but different to the isotropy group SO(3) of the massive particle. The general state is given by equation (4.1.8) provided we take $\varphi_m= 0=\bar \varphi_{m}$ for $|m|\ge 2$. It therefore depends on $\phi$, $\bar \phi$, $\chi$,  $\bar \chi$ and $\chi_0$. As such the  infinite number of super momenta satisfy an infinite number of constraints.  
\par
The {\bf massless} irreducible representations of $BMS_4$ were given in section (4.2) and  taking only  parameters that correspond to the Lorentz group we find that  $\Lambda_{\pm 1}=0$ and  $\bar \Lambda_{\pm 1}=0$ leaving only $\Lambda_0$ and $\bar \Lambda_0$ which obey the condition $\Lambda_0+\bar\Lambda_0=0$. Thus the isotropy group has only one generator $J_0-\bar J_0$ which is the same as when the super rotations were present and in effect agrees with that of the massless Poincar\'e particle when we take 
account of the fact that $J_{1+}$ and $J_{2+}$ are trivially realised. The general state is given by equation (4.2.4) provided we set $\varphi_m$ and $\varphi_m$ to zero for $|m|\ge 2$. Again  the momenta obey an infinite number of constraints. 
\par
For the orbits with the same representative as in the Poincar\'e case that we are considering here, the irreducible representations of BMS, when super rotations are absent, are much more trivial than when super rotations are present, since the boosts contain only the generators of the Lorentz group that lie outside the isotropy group. As a result even though there are an infinite number of super momenta they are trivial in the sense that they satisfy an infinite number of constraints that express them in terms of the usual momenta. This is a general feature of representations of BMS when the super rotations are absent. 
\par 
Above we truncated the super rotations in the extended $BMS_4$ algebra and found the irreducible representations. However this is not the same as the global $BMS_4$ algebra of Bondi, Metzner and Sachs  that also has just Lorentz rotations,  but using a spherical harmonic decomposition for the super translations. This algebra has momenta 
$Z_{j,m}$ where $j=0, 1, 2, \ldots $ , $m\le |j|$ and it is given by [38]
$$
[J_m, J_n] = (m-n) J_{m+n}
\qquad
[\bar J_m, \bar J_n] = (m-n) \bar J_{m+n}
\qquad
[J_m, \bar J_n] = 0 
\eqno(5.1.5)$$
 $$
[J_{-1}, Z_{j,m}]
= {(j+2)(j+m)(j+m-1)\over 4(2j+1)(2j-1)}\, Z_{j-1,m-1}
- {j+m\over 2}\, Z_{j,m-1}
+ (j-1)\, Z_{j+1,m-1} 
$$
$$
[J_0, Z_{j,m}]
= -{(j+2)(j+m)(j-m)\over 4(2j+1)(2j-1)}\, Z_{j-1,m}
- {m\over 2}\, Z_{j,m}
+ (j-1)\, Z_{j+1,m} 
$$
$$
[J_1, Z_{j,m}]
= {(j+2)(j-m)(j-m-1)\over 4(2j+1)(2j-1)}\, Z_{j-1,m+1}
+ {j-m\over 2}\, Z_{j,m+1}
+ (j-1)\, Z_{j+1,m+1} 
 $$
$$
[\bar J_{-1}, Z_{j,m}]
= -{(j+2)(j-m)(j-m-1)\over 4(2j+1)(2j-1)}\, Z_{j-1,m+1}
+ {j-m\over 2}\, Z_{j,m+1}
- (j-1)\, Z_{j+1,m+1} 
$$
$$
[\bar J_0, Z_{j,m}]
= -{(j+2)(j+m)(j-m)\over 4(2j+1)(2j-1)}\, Z_{j-1,m}
+ {m\over 2}\, Z_{j,m}
+ (j-1)\, Z_{j+1,m} 
 $$
$$
[\bar J_1, Z_{j,m}]
= -{(j+2)(j+m)(j+m-1)\over 4(2j+1)(2j-1)}\, Z_{j-1,m-1}
- {j+m\over 2}\, Z_{j,m-1}
- (j-1)\, Z_{j+1,m-1} 
$$
$$
[Z_{j,m}, Z_{j',m'}] = 0 
 \eqno(5.1.6)$$
We identify the usual Poincar\'e algebra as the identification of equation (4.0.3) as well as 
$$
{\bf P}_0 = Z_{0,0}, \qquad {\bf P}_1 = Z_{1,1} - Z_{1,-1}, \qquad {\bf P}_2 = i\left(Z_{1,1} +Z_{1,-1}\right), \qquad {\bf P}_3 = -2 Z_{1,0}.
\eqno(5.1.7)$$
\par
In the {\bf massive}  case we take ${\bf p}_0^{(0)}=z_{0,0}^{(0)}=m$ where we  have used the symbol $z_{j,m}$ to denote the momenta in  this basis. Inserting this value on the right-hand side of the commutators of equation (5.1.6) we find that this choice is preserved if 
$$
{\delta z_{1,1}\over m}= {1\over 2} (\Lambda_1-\bar \Lambda_{-1})=0, \  {\delta z_{1,0}\over m}= -{1\over 4} (\Lambda_{0}+\bar \Lambda_{0})=0,
{\delta z_{1,-1}\over m}= {1\over 2} (\Lambda_{-1}-\bar \Lambda_{1})=0, \ 
\eqno(5.1.8)$$
all other variations being trivially satisfied. Thus we find the conditions of equation (4.1.4) and so  the isotropy group SO(3), 
which is the same isotropy group as the massive irreducible representation as the Poincar\'e group but not the same as for extended $BMS_4$ which was SO(2). The general state has the form of equation (4.1.10) provided we set $\varphi_n=0$. 
\par
Now we will find the  {\bf massless} irreducible representations for which we take 
$p_0^{(0)}=-{1\over \sqrt 2}=-p_3^{(0)}$, or equivalently $z^{(0)}_{1,0}= -{1\over 2\sqrt 2}$ and $z_{0,0}^{(0)}= -{1\over \sqrt 2}$. One finds, using commutators of equation (5.1.6), that the only non-trivial results are 
$$
\sqrt {2}\delta z_{1,1}= \bar \Lambda_{-1}=0, \  \sqrt {2}\delta z_{1,0}= {1\over 4} (\Lambda_{0}+\bar \Lambda_{0})=0,
\sqrt {2}\delta z_{1,-1}= -\Lambda_{-1}=0,
$$
$$
 \sqrt {2}\delta z_{0,0}= {1\over 2} (\Lambda_{0}+\bar \Lambda_{0})=0
\eqno(5.1.9)$$
as well as 
$$
\sqrt {2}\delta z_{2,1}= -{1\over 5}(\Lambda_{1}-\bar \Lambda_{-1})=0, \ 
\sqrt {2}\delta z_{2,-1}= -{1\over 5}(\Lambda_{-1}-\bar \Lambda_{1})=0, \ 
$$
$$
\sqrt {2}\delta z_{2,0}= {4\over 15} (\Lambda_{0}+\bar \Lambda_{0})=0
\eqno(5.1.10)$$
The conditions of equation (5.1.9) are the same as those of the massless Poincar\'e particle, as must be the case,  but the variation of the super momentum leads to the  additional conditions of equation (5.1.10) that leave only $\Lambda _0$ and  $\bar \Lambda _0$
subject to $\Lambda_{0}+\bar \Lambda_{0}=0$. As such we have the same isotropy group as for extended $BMS_4$. We recall that in the Poincar\' e case the isotropy group is,  in effect,  reduced to SO(2) and so to this extent it is the same. 
\par
Since we only have at most three and five boosts for the massive and massless cases  the general state of the massive and massless irreducible representations respectively,  the super momenta satisfy an infinite number of constraints as they did for global $BMS_4$. As such these representations are rather trivial compared to extended $BMS_4$.
\par
The little groups discussed above for the massive and massless cases all appear in the classification of McCarthy [14], and one can check that the corresponding Poincar\'e representatives considered here are in agreement with those discussed in that reference. It is worth mentioning that additional little groups have been discussed later, for example in [32], and arise in a different functional-analytic setting, namely by changing the topology on the supertranslation space and allowing distributional representatives. In particular, the $ISO(2)$ little group, which is the same as the massless Poincar\'e little group before imposing additional constraints, can be obtained in that way. 


\medskip
{\bf 6. Summary and Discussion}
\medskip
In this paper we have constructed the irreducible representations of the BMS group in three and four dimensions that correspond to the  massive and massless irreducible representations of the Poincar\'e group, or put another way the Poincar\'e particle. More precisely we took the BMS representations to  have the same rest frame momenta as the Poincar\'e particles.  We found that 

$ \bullet$ The massive irreducible representation of  $BMS_3$ has the isotropy group SO(2). This  agrees with that of the massive  Poincar\'e particle in three dimensions and it contains this representation.  While if we included a central charge with a very precise value the isotropy group was enhanced to SL(2,R). These isotropy groups agree with those found in references [18,19,20]

$ \bullet$ The massless  irreducible representation of $BMS_3$ has an isotropy group that has  two generators  even if the central charge $c_2$ is non-zero. This contrasts with that of the massless Poincar\'e particle in three dimensions, whose isotropy group has only one generator. However, the commutator of the two generators is ill defined and so it is difficult to make a comparison with the massless Poincar\'e particle. To properly understand this result requires further study. 

$ \bullet$ The massive irreducible representation of extended $BMS_4$ has isotropy group SO(2) while the isotropy group of the massive Poincar\'e particle in four dimensions  is SO(3). As such the former does not contain the latter representation,  but we showed  that there is a reducible representation of extended $BMS_4$ that does contain that of the massive Poincar\'e particle. 

$ \bullet$ The massless irreducible representation of extended $BMS_4$ has isotropy group SO(2) while the isotropy group of the irreducible massless  Poincar\'e particle in four dimensions  is $SO(2)\otimes_s T^2$. However for the usual representations of the massless Poincar\'e particle the generators of the $T^2$, that is $J_{1+}$ and  $J_{2+}$, are trivially realised and in this case the isotropy groups coincide and the former representations contains the latter. 

The isotropy group of representations of BMS is generally smaller than that of the Poincar\'e group as the variation of some of the higher momenta, the super momenta,  can transform  under a Lorentz transformation into a non-zero frame momenta. 
\par
The relationship between the irreducible representation of extended BMS algebra and its Poincar\'e subalgebra is more complicated than one might naively have expected. However, one can always  find a representation, perhaps reducible, of the former that contains the usual massive and massless irreducible representations of the latter. 
 \par
 We also constructed the equivalent irreducible representations of  $BMS_3$ and extended $BMS_4$ in the absence of the super rotations. We found the super momenta are subject to an infinite number of constraints to leave in effect only the momenta of the Poincar\'e group and as such the representations are rather trivial compared to those when the super rotations are present. 
\par
In this paper we have constructed the above irreducible representations in detail  paying particular attention to the role of the super rotations. The general state is found by boosting the  rest frame state $|p^{(0)}\rangle$ by those rotations that are not in the isotropy group. As the isotropy group is very small this contains  an infinite number of generators which are of the generic  form $e^{\varphi\cdot J}$. The momenta of the general state is determined in terms of the $\varphi$ and so the number of independent momenta is the same as the number of $\varphi$. As such the number of constraints the momenta satisfy is the difference in the number of momenta and the number of $\varphi$. 
\par
This is just the same as for the Poincar\'e case,  for example for  a massive particle  in four dimensions there are three boosts (${\bf J}_{0i}, i=1,2,3$) and so three $\varphi$  leading to only three independent momenta. Put another way ${\bf p}_{\mu}$ is subject to the one constraint  ${\bf p}^\mu {\bf p}_\mu+m^2=0$.
\par
To find the wave function in position space we took  the Fourier transform of the general state of the irreducible representation with a factor $e^{ip\cdot x}$ and integrate over the independent components of the momenta, or equivalently the $\varphi$. To do this one must introduce one position space coordinate for every independent momentum. Since there are an infinite number of the latter we must introduce an infinite number of position coordinates. As the BMS symmetries were deduced in the usual spacetime it is not immediately clear what is the meaning of these infinite number of additional coordinates. 
\par
We showed that the position space coordinates can generically be encoded in an object  $X(z)$ which at its lowest level contains the usual coordinates of spacetime. The BMS symmetry is then realised as a reparameterisation of $z$. As such in position space the irreducible representations of BMS are carried by functionals of $X(z)$, that is $\Psi (X(z))$. Thus one can think of this as a string with parameter $z$. Although this result has been discussed in the context of the representations discussed in this paper,  it is very likely to apply to any representations in which the isotropy group is of finite dimension as in this case the boost has   an infinite number of parameters  and so there are an infinite number of momenta and as a result an infinite number of position coordinates. 
\par
It would be interesting to find the dynamics that this string obeys. The dynamics of this string is $BMS_3$ invariant and so  it  must be invariant under the reparameterisation of the world line of equation (2.1.20) which is that for an object with an upper world index. 
In a Hamiltonian approach one would expect that such a local symmetry would lead to first class constraints and it would be interesting to see how these are realised on the string states. 
\par
We showed in section (3.1) that the irreducible representation for the massive Poincar\'e particle can be described in position space by an unconstrained  function  at time-like infinity $i^+$ which carried a representation of the isotropy group. We then argue that  this result can be generalised to  representations of extended BMS in three and four dimensions in that these can be described by a string, more precisely a functional of a string living on $i^+$. 
\par
One obvious way to extend the results of this paper is to compute the extended BMS variations of the general state of the representation as was done, for example in reference [34] for the massless Poincar\'e particle. One could  also try to find a covariant representation as one does for the Poincar\'e group. In this regard it would be interesting to find what equations of motion the irreducible representations satisfy and in particular what equations the string fields satisfy. The latter would be strongly constrained by the BMS symmetries which manifest themselves as reparameterisations. Once one had the equation of motion one could quantise the string and it would be very interesting to know what is the physical meaning of the quantum excitations and if they had something to do with the infrared effects and, in particular,  particle dressing.  
\par
That a duality connects objects of different dimensions is not new as the AdS/CFT correspondence relates strings to point particles. It would be of interest to see what is the significance of the strings we have identified in the context of the  Carrollian proposal for flat space holography. It would seem that the scattering of particles in Minkowski space should be related to the scattering of strings at infinity. It would also be interesting to investigate if there is a connection to the work of references [36] and [37] which consider a world sheet action at null infinity. This construction is based on the ambitwistor string [39] which is related to the null string obtained by taking the tensionless limit of a relativistic string [40,41,42].  

\medskip
{\bf Acknowledgement} 
\medskip
We would like to thank Glenn Barnich for discussions and, in particular,  on the role of topology of the super momenta in determining the isotropy group. We also thank Blagoje Oblak for a number of very helpful comments  on the first version of this paper and in particular for drawing our attention to the role played by the central charge in representation theory in three dimensions. 
 RR is supported by the European Union's Horizon Europe research and innovation programme under the Marie Sklodowska-Curie grant agreement No. 101104845 (UniFlatHolo), hosted at Harvard University and Ecole Polytechnique. PW would like to thank STFC research council for support over many years. 

\medskip
{\bf References}
\medskip

\item{[1]} H. Bondi, M. van der Burg and A. Metzner, {\it Gravitational waves in general relativity. 7. Waves from axisymmetric isolated systems}, Proc. Roy. Soc. Lond. A (1962), 269, 21�.

\item{[2]} R. Sachs, {\it Asymptotic symmetries in gravitational theory}, Phys. Rev. (1962), 128, 2851�64.

\item{[3]} G. Barnich and C. Troessaert, {\it Symmetries of asymptotically flat 4 dimensional spacetimes at null infinity revisited}, Phys. Rev. Lett. (2010), 105, 111103, arXiv:0909.2617.

\item{[4]} G. Barnich and C. Troessaert, {\it Supertranslations call for superrotations}, PoS (2010), CNCFG2010, 010, arXiv:1102.4632.

\item{[5]} G. Barnich and C. Troessaert, {\it Aspects of the BMS/CFT correspondence}, JHEP (2010), 05, 062, arXiv:1001.1541.

\item{[6]} G. Barnich and C. Troessaert., {\it BMS charge algebra}, JHEP (2011), 12, 105, \hfil\break arXiv:1106.0213.

\item{[7]} M. Campiglia and A. Laddha, {\it Asymptotic symmetries and subleading soft graviton theorem}, Phys. Rev. D (2014), 90(12), 124028, arXiv:1408.2228.

\item{[8]} M. Campiglia and A. Laddha, {\it New symmetries for the Gravitational S-matrix}, JHEP (2015), 04, 076, arXiv:1502.02318.

\item{[9]} G. Comp{\`e}re, A.  Fiorucci  and R. Ruzziconi, {\it Superboost transitions, refraction memory and super-Lorentz charge algebra}, JHEP (2018), 11, 200, arXiv:1810.00377 [Erratum: JHEP 04, 172 (2020)].

\item{[10]} E. Flanagan, K.  Prabhu and I. Shehzad, {\it Extensions of the asymptotic symmetry algebra of general relativity}, JHEP (2020), 01, 002, arXiv:1910.04557.

\item{[11]} M. Campiglia and J.  Peraza, {\it Generalized BMS charge algebra}, Phys. Rev. D (2020), 101(10), 104039, arXiv:2002.06691.

\item{[12]} A. Strominger., {\it Lectures on the Infrared Structure of Gravity and Gauge Theory}, Princeton University Press (2018), arXiv:1703.05448.

\item{[13]} P. McCarthy., {\it Asymptotically flat space-times and elementary particles}, Phys. Rev. Lett. (1972), 29, 817�9.

\item{[14]}P. McCarthy, {\it Representations of the Bondi Metzner Sachs group I. Determination of the representations}, Proc. Roy. Soc. Lond. A (1972), 330, 517.

\item{[15]} P. McCarthy, {\it Representations of the Bondi Metzner Sachs Group. II. Properties and Classification of the Representations}, Proc. Roy. Soc. Lond. A (1973), 333, 317�6.

\item{[16]} P. McCarthy and  M. Crampin, {\it Representations of the Bondi Metzner Sachs group. III. Poincare spin multiplicities and irreducibility}, Proc. Roy. Soc. Lond. A (1973), 335, 301�1.

\item{[17]} P. McCarthy and M. Crampin, {\it Representations of the Bondi Metzner Sachs group. IV. Cantoni representations are induced}, Proc. Roy. Soc. Lond. A (1976), 351, 55�.

\item{[18]} G. Barnich and B.  Oblak, {\it Notes on the BMS group in three dimensions: I. Induced representations}, JHEP (2014), 06, 129, arXiv:1403.5803.

\item{[19]} G. Barnich and B. Oblak, {\it Notes on the BMS group in three dimensions: II. Coadjoint representation}, JHEP (2015), 03, 033, arXiv:1502.00010.

\item{[20]} B. Oblak, {\it BMS Particles in Three Dimensions}, PhD Thesis, U. Brussels (2016), arXiv:1610.08526.

\item{[21]} X. Bekaert, L. Donnay and Y. Herfray, {\it BMS particles}, Phys. Rev. Lett. (2025), 135(13), 131602, arXiv:2412.06002.

\item{[22]} X. Bekaert and Y. Herfray, {\it BMS representations for generic supermomentum}, \hfil\break arXiv:2505.05368 (2025).

\item{[23]} G. Arcioni and C. Dappiaggi, {\it Exploring the holographic principle in asymptotically flat space-times via the BMS group} , 
Nucl. Phys. {\bf B {674}} (2003) 553-592, arXiv:hep-th/0306142 [hep-th]].
\item{[24]} G. Barnich, {\it Centrally extended BMS4 Lie algebroid},  JHEP {\bf 06} (2017) 007, \hfil\break arXiv:1703.08704 [hep-th].
\item{[25]} C. Duval, G. Gibbons and P. Horvathy, {\it Conformal Carroll groups and BMS symmetry},' Class. Quant. Grav. {\bf 31} (2014), 092001, [arXiv:1402.5894 [gr-qc]].
\item{[26]} A. Bagchi, R. Basu, A. Kakkar and A. Mehra, {\it Flat Holography: Aspects of the dual field theory}, JHEP {\bf 12} (2016) 147, [arXiv:1609.06203 [hep-th]].
\item{[27]} L Ciambelli, C Marteau, A Petkou, P Petropoulos and K Siampos, {\it Flat holography and Carrollian fluids}, 
JHEP {\bf 07} (2018), 165, [arXiv:1802.06809 [hep-th]].
\item{[28]} L Donnay, A Fiorucci, Y Herfray and R Ruzziconi, {\it Carrollian Perspective on Celestial Holography},'
Phys. Rev. Lett. {\bf 129} (2022) no.7, 071602, [arXiv:2202.04702 [hep-th]].
\item{[29]} L Donnay, A Fiorucci, Y Herfray and R Ruzziconi, {\it Bridging Carrollian and celestial holography}, Phys. Rev. D {\bf 107} (2023) no.12, 126027, PhysRev D.107.126027, [arXiv:2212.12553 [hep-th]].
\item{[30]} Wigner, E. P., {\it On Unitary Representations of the Inhomogeneous Lorentz Group}, Annals Math. (1939), 40, 149�4.
\item{[31]} A Ashtekar, J Bicak and B Schmidt,{\it  Asymptotic structure of symmetry reduced general relativity}, Phys. Rev. {\bf D 55} (1997) 669-686, [arXiv:gr-qc/9608042 [gr-qc]].

\item{[32]}P. J. McCarthy, {\it The Bondi-Metzner-Sachs group in the nuclear topology}, Proc. Roy. Soc. Lond. A (1975), 343, 489--523.

\item{[33]} G Barnich and G Compere, {\it Classical central extension for asymptotic symmetries at null infinity in three spacetime dimensions}, Class. Quant. Grav. {\bf 24} (2007), F15-F23
\item{[34]} K. Nguyen and P. West, {\it  Carrollian Conformal Fields and Flat Holography},  Universe {\bf 9} (2023) 385, [arXiv: 2305.02884 [hep-th]
\item{[35]} E. Have, K. Nguyen, S. Prohazka and J. Salzer, {\it Massive carrollian fields at timelike infinity},     JHEP 07 (2024) 054, 
 [arXiv: 2402.05190 [hep-th]
 \item{[36]} T Adamo, E Casali and D Skinner, {\it Perturbative gravity at null infinity},'Class. Quant. Grav. {\bf 31} (2014) no.22, 225008, [arXiv:1405.5122 [hep-th]].
\item{[37]} T Adamo and E Casali, {\it Perturbative gauge theory at null infinity},' Phys. Rev. {\bf D91} (2015) no.12, 125022, 
[arXiv:1504.02304 [hep-th]].
\item{[38]} G. Barnich and R. Ruzziconi , {\it Coadjoint representation of the BMS group on celestial
Riemann surfaces}, JHEP {\bf 06} (2021), 079, [arXiv: 2103.11253  [gr-qc]
\item{[39]} L Mason and D Skinner, {\it Ambitwistor strings and the scattering equations},  JHEP {\bf 07} (2014), 048, arXiv:1311.2564 [hep-th].

\item{[40]} A Schild, {\it Classical Null Strings},' Phys. Rev. {\bf D16} (1977) 1722

\item{[41]} J Isberg, U Lindstrom, B Sundborg and G Theodoridis, {\it Classical and quantized tensionless strings}, 
Nucl. Phys. {\bf B 411} (1994), 122, [arXiv:hep-th/9307108 [hep-th].

\item{[42]}   A Bagchi, S Chakrabortty and P Parekh, {\it Tensionless Strings from Worldsheet Symmetries}, 
JHEP {\bf 01} (2016), 158, arXiv:1507.04361 [hep-th].

\item{[43]}   A Bagchi, A Banerjee, S Chakrabortty, S Dutta and P Parekh, {\it A tale of three  tensionless strings and vacuum structure}, JHEP {\bf 04} (2020), 061, arXiv:2001.00354 [hep-th].

\item{[44]} W Donnelly, L Freidel, S Moosavian and A Speranza, {\it Gravitational edge modes, coadjoint orbits, and hydrodynamics},'
JHEP {\bf 09} (2021) 008, arXiv:2012.10367 [hep-th].

\item{[45]} D Kapec, V Lysov, S Pasterski and A Strominger, {\it Semiclassical Virasoro symmetry of the quantum gravity l{\bf S}-matrix} , JHEP {\bf 08} (2014) 058, arXiv:1406.3312 [hep-th].

\item{[46]} A. Campoleoni, H. Gonzalez, B. Oblak and M. Riegler, {\it BMS Modules in Three Dimensions},�Int. J. Mod. Phys. A {\bf 31} (2016) no.12, 1650068, [arXiv:1603.03812 [hep-th]].

\item{[47]}  M. Campiglia and A. Laddha, {\it Asymptotic symmetries of gravity and soft theorems for massive particles}, JHEP {\bf 12} (2015), 094, [arXiv:1509.01406 [hep-th]].

\end